\documentclass{aa}  
\usepackage{amsmath}
\usepackage{ amssymb}
\usepackage{graphicx}
\usepackage{graphics}
\usepackage{subfigure}
\usepackage{color}
\usepackage{txfonts}
\usepackage{natbib}
\bibpunct{(}{)}{;}{a}{}{,}

\def\teff{\ifmmode T_{\rm eff} \else $T_{\mathrm{eff}}$\fi}

\def\ltsima{$\buildrel<\over\sim$}
\def\lsim{\lower.5ex\hbox{\ltsima}}
\def\nh{$N_{\rm HI}$}

\def\speak{$S_{\rm peak}$}
\def\fwhm{$FWHM$}
\def\ew{EW}
\def\ebv{$E(B-V)$}

\newcommand{\hi}{H{\sc i}}
\newcommand{\hii}{H{\sc ii}}
\newcommand{\ha}{\ifmmode {\rm H}\alpha \else H$\alpha$\fi}
\newcommand{\hb}{\ifmmode {\rm H}\beta \else H$\beta$\fi}
\newcommand{\lya}{\ifmmode {\rm Ly}\alpha \else Ly$\alpha$\fi}

\newcommand{\McLya}{{\tt McLya}}
\newcommand{\vexp}{\ifmmode v_{\rm exp} \else $v_{\rm exp}$\fi}
\newcommand{\vpeak}{\ifmmode v_{\rm peak} \else $v_{\rm peak}$\fi}

\def\fesclyc{$f_{\mathrm{esc}}^{\mathrm{LyC}}$}

\def\kms{km s$^{-1}$}

\def\hii{\ion{H}{ii}}
\def\oii{[\ion{O}{ii}]}

\def\oiill{[\ion{O}{ii}]$\lambda\lambda 3726,3729$}
\def\oiii{[\ion{O}{iii}]}
\def\oiiil{[\ion{O}{iii}]$\lambda 5007$}
\def\oiiill{[\ion{O}{iii}]$\lambda\lambda 4959,5007$}

\begin{document}

\title{Using Lyman-$\alpha$ to detect galaxies that leak Lyman continuum}

\author{Anne Verhamme\inst{1}, Ivana Orlitov\'a\inst{1,2}, Daniel Schaerer\inst{1,3}, Matthew Hayes\inst{4} . 
}
\offprints{anne.verhamme@unige.ch}
\institute{
Observatoire de Gen\`eve, Universit\'e de Gen\`eve, 
51 Ch. des Maillettes, 1290 Versoix, Switzerland
\and
Astronomical Institute, Academy of Sciences of the Czech Republic, 
Bo\v cn{\'\i} II 1401, 141 00 Prague, Czech Republic
\and
CNRS, IRAP, 14 Avenue E. Belin, 31400 Toulouse, France
\and
Department of Astronomy, Oskar Klein Centre, 
Stockholm University, AlbaNova University Centre, 
SE-106 91 Stockholm, Sweden
}
\date{Received date / Accepted date}
\authorrunning{Verhamme et al.}{ }
\titlerunning{\lya\ as a proxy for LyC?}{ }

\abstract{}
{ We propose to infer the output of the ionising continuum-leaking properties of galaxies
based upon their \lya\ line profiles.
} 
{We carried out \lya\ radiation transfer calculations in two models of 
\hii\ regions. These models are porous to ionising continuum escape:
1) we define Lyman-continuum (LyC) optically thin star clusters, in which massive stars
produce enough ionising photons to keep the surrounding interstellar medium
transparent to the ionising continuum, in other words, almost totally ionised, and
2) we define riddled ionisation-bounded media that are 
surrounded by neutral interstellar medium,
but have holes, which results in a covering fraction lower than unity.
}
{The \lya\ spectra that emerge from these configurations have distinctive
features:
1) a classical asymmetric redshifted profile in the first case, but 
with a small shift of the profile maximum compared to the systemic 
redshift (\vpeak\ $\leq 150$\,\kms);
2) a main peak at the systemic redshift in the second case (\vpeak\,$= 0$),
with a non-zero \lya\ flux bluewards of the systemic redshift as a consequence.
If in a galaxy that leaks ionising photons the \lya\ component that emerges from the
leaking star cluster(s) is assumed to dominate the total \lya\ spectrum,  
the \lya\ shape may be used as a pre-selection tool for detecting LyC-leaking galaxies in objects with high spectral resolution \lya\ spectra (R $\geq 4000$).
Our predictions are corroborated by examination of a sample of ten local starbursts with high-resolution HST/COS 
\lya\ spectra that are known in the literature as LyC leakers 
or leaking candidates.
} 
{Observations of \lya\ profiles at high resolution are expected
to show definite signatures 
revealing the escape of Lyman-continuum photons from star-forming galaxies.}

\keywords{Radiative transfer -- Reionisation -- Galaxies: ISM -- ISM: structure, 
kinematics and dynamics}

\maketitle

\section{Introduction}
\label{s_intro}

Determining the population of objects that reionised the Universe and
maintained the subsequent thermal history of the intergalactic medium
(IGM) remains an outstanding and urgent question in observational
cosmology.  Observations show that some active galactic
nuclei were in place almost at the reionisation epoch, but their numbers
were grossly insufficient to reproduce the necessary background of Lyman-continuum (LyC; $<912$\AA) radiation \citep[e.g.][]{Cowie09, Fontanot12}.
This leaves star-forming galaxies to reionise the Universe, and, while
they undoubtedly are in place at the relevant epoch, with hints for a steep
luminosity function (LF) at the faint end \citep{Alavi14}, it still
appears that the current populations are insufficient to have completed
reionisation by $z\approx 6$ \citep[e.g.][and references therein]{Robertson13}.

Moreover, simply knowing that galaxies are in place is not sufficient:
they must also emit enough LyC radiation.  In the very nearby
Universe this is manifestly not the case, where space-borne UV
telescopes such as the HUT (Hopkins Ultraviolet Telescope) 
and FUSE (Far Ultraviolet Spectroscopic Explorer) have reported a large number of
upper limits \citep[e.g.][]{Leitherer95, Heckman01, Deharveng01, Grimes09}
and only a small number of weak detections with escape fractions of only a
few percent \citep[e.g.][]{Leitet11,Leitet13, Borthakur14}.
Despite the much larger samples of galaxies that have been studied in
LyC at $z\sim 1$ with the Galaxy Evolution Explorer (GALEX) and 
the Hubble Space Telescope (HST), the situation still does not change
much, and still no individual LyC-leaking galaxies are reported
\citep{Malkan03, Siana07, Cowie09, Siana10}.

The beginnings of the necessary evolution seem to set in at higher
redshifts of $z\sim 3$.  LyC emission has been reported in
large samples of Lyman break galaxies (LBGs) and \lya-emitters (LAEs),
using both spectroscopic \citep{Steidel01, Shapley06} techniques and
narrowband imaging bluewards of the restframe Lyman limit
\citep{Iwata09, Nestor11, Nestor13, Mostardi13}.  Curiously, most
narrowband images of the emitted LyC
show spatial offsets from the non-ionising UV continuum, which may
indeed be consistent with models of LyC leakage being facilitated by supernova
winds (Clarke \& Oey 2002) or unresolved galaxy mergers (Gnedin et al.
2008).  Alternatively, the spatial offsets may also be explained by projected
galaxies at lower redshift that contaminate the observed LyC \citep{Vanzella10,
Vanzella12}, although simulations performed by \citet{Nestor13} suggest
that this probably does not account for all detections.

Another particularly relevant result from $z\sim 3$ is that LyC leakage
seems to be stronger from LAEs than from LBGs.  Theorists have
repeatedly shown that LyC leakage generally increases at lower galaxy
masses \citep[e.g.][]{Ferrara13, Yajima11, Wise09}, which superficially
seems to be consitent with the stronger LyC leakage found from $z\sim 3$
LAEs, which tend to occupy lower mass haloes \citep{Ouchi05, Gawiser06,
Guaita11}.  Because the UV LFs found in the high-$z$ Universe
are particularly steep \citep{Alavi14, Dressler14} and LyC escape fractions increase 
moving downwards across the galaxy LF, the case may indeed be that 
faint \lya-emitting galaxies were the main contributors to reionisation.
The evolution of the number of \lya\ emitting galaxies among 
the population of LBGs with a sudden drop at $z > 6$ \citep{Stark10, Ono12, Schenker14}
is usually interpreted as due to an increase of the neutral fraction of the IGM.
But if \lya-emitting galaxies have a non-negligible LyC 
escape fraction ($\geq 15\%$) at z>6 and if this increases again towards higher z, 
then these galaxies become fainter in \lya\ at levels that could (partially) mimic 
a reionisation signature \citep{Hayes11, Dijkstra14}.
This reasoning can be generalised to all nebular lines: 
along the same idea, \citet{Zackrisson13} proposed to search for weak nebular lines 
in the spectra of galaxies of the reionisation epoch as a probe for LyC leakage. 
In the local Universe, \citet{Lee09} observed a systematic underestimate of the 
star formation rate derived from H$\alpha$ luminosity, SFR(H$\alpha$), 
compared to SFR(UV), the star formation rate derived from ultraviolet luminosity, 
in dwarf galaxies among their sample of nearby galaxies, 
which can be interpreted as the result of a higher escape of ionising photons 
in smaller galaxies.

To explain the low success rate of LyC-leaking detections, 
the commonly invoked explanation is that the galaxies responsible for reionisation are
the faintest ones, which are below our current continuum detection threshold \citep[e.g.][]{Ouchi08}.
On the other hand, a huge amount of spectroscopic data are available in \lya, 
up to the reionisation redshift 
\citep[e.g.][]{Shapley03,Hu10,Stark10,Guaita10,Guaita11,Dressler11,Bielby11,
Pentericci10,Pentericci11, Kulas12,Jiang13a, Jiang13b, Ellis13, Schenker13}. 
Could we tell from the \lya\ line shape if a galaxy is a good candidate for 
continuum leaking? This is the starting point of our study.
Based on \lya\ radiation transfer modelling, 
we first present the theoretically expected \lya\ spectral characteristics of 
continuum-leaking galaxies.
We then compare our diagnostics with two 
other indirect diagnostics of continuum leaking that  have been
proposed in the literature, which are 
(a) a low covering fraction of the interstellar absorption lines
that are in a  low-ionisation state (LIS)\citep[e.g.][]{Heckman11, Jones13}, and 
(b) a high ratio of \oiiil/\oiill\  as a proxy for density-bounded \hii\ regions
\citep{Nakajima14,Jaskot13, Kewley13}.
Finally we compare the different indicators for a sample of low-redshift galaxies.

Our paper is structured as follows. The link between Lyman-continuum leakage and
the \lya\ line profile is discussed in Sect.\ \ref{s_LyaLyC}.
In Sect.\ \ref{s_compare} we compare the different leaking indicators 
for a sample of low-redshift galaxies.
Our main results are discussed in Sect.~\ref{s_discuss} and 
are summarised in Sect.\ \ref{s_conclude}.

\section{Link between LyC and \lya}
\label{s_LyaLyC}

In this section, we describe the possible implications that ionising continuum 
leakage from a galaxy may have on its \lya\ profile, and we discuss the detectability 
of these spectral features.

\subsection{Ionising continuum leakage from galaxies}

As described in the introduction, galaxies from which ionising photons escape 
are extremely rare in the local Universe, at least at the observed luminosities.
Only three objects are known as LyC leakers, and their absolute escape 
fraction\footnote{$f_{esc} = f_{esc,rel}\times 10^{-0.4\times A_{1500}} = 
\frac{(f_{1500}/f_{900})_{int}}{(f_{1500}/f_{900})_{obs}}\times 10^{-0.4\times A_{1500}}$, 
as defined in \citet{Leitet13}} is only a few percent \citep{Bergvall06,Leitet13,Borthakur14}. 
The vast majority of known high-redshift galaxies are also opaque to their own 
ionising radiation, as demonstrated by the success of the well-known Lyman-break 
selection \citep[as emphasised recently in][]{Cooke14}. 
Only a few tens of objects are detected in the LyC 
\citep{Iwata09, Nestor11, Nestor13, Mostardi13}.
A LyC-leaking galaxy has then peculiar rare properties that we describe below.

\subsubsection{Two mechanisms of ionising continuum escape}

Our framework is the following: we consider that the bulk of ionising 
photons produced in galaxies arises from young massive stars, 
and we neglect AGN or external sources (proximate quasar, UV background, etc.).
Young star clusters ionise the surrounding ISM, creating \hii\ regions 
that are usually ionisation bounded, when the size of their birth 
gas cloud exceeds the Str\"omgren radius \citep{Stromgren39}, 
but that can be density bounded in peculiar cases 
when their surrounding cloud is not large enough to absorb all 
the ionising radiation \citep[e.g.][]{Pellegrini12}. 
Density-bounded \hii\ regions are leaking LyC photons by definition,
with an escape fraction of
\begin{equation}
\label{eq_fesc}
f_{\rm esc}(LyC) = e^{-\tau_{\rm ion}},
\end{equation}
where $\tau_{\rm ion} = \sigma(\nu) N_{\rm HI}$ is the optical depth 
seen by an ionising photon of frequency $\nu$, $N_{\rm HI}$ is the
column density of the remaining neutral gas surrounding the star cluster, and 
\begin{equation}
\sigma(\nu) = \sigma_{\nu_0} (\nu / \nu_0)^{-3}  
\end{equation}
is the photoionisation cross-section of a hydrogen atom,
defined for $\nu \geqslant \nu_0$, where $\nu_0$ is the ionisation 
frequency of the hydrogen atom for which the ionisation 
cross-section is  $\sigma_{\nu_0} = 6.3\times10^{-18}$ cm$^2$.

Usually, and for the remainder of this paper, \fesclyc\ is considered at the Lyman edge, 
that is,\ for $\nu \approx \nu_0$. 
This is also the range probed by observations using 
narrowband (NB) technics at most redshifts.
 
In addition to the optically thin regime, 
we consider another scenario for LyC leaking from galaxies
\citep[following][]{Zackrisson13}:
the ionising continuum produced by young 
massive stars in an ionisation-bounded region ($\tau_{\rm ion} \gg 1$)
can escape in some peculiar cases where the covering fraction of the neutral 
gas surrounding the cluster is not unity. We call this second configuration of LyC 
escape riddled ionisation-bounded \hii\ regions.
The covering fraction, CF, is defined as the fraction of lines of sight 
that are optically thick to LyC photons. This implies that some lines of sight are 
transparent to LyC photons and some are not \citep[e.g.][]{Behrens14}. 
The escape fraction of the ionising radiation
is simply related to the covering fraction of the neutral gas: 
\begin{equation}
f_{\rm esc}(LyC) = 1 - CF.
\end{equation}

\subsubsection{Link with the \lya\ radiation}

\lya\ radiation is produced by the recombination of hydrogen atoms 
in these \hii\ regions that surround young star clusters. 
The quantity of \lya\ photons produced in a galaxy is then proportional to  
the non-escaping fraction of ionising photons (1 - f$_{\rm esc}$(LyC)).
Almost all LyC leakers that have been reported in the literature 
have an escape fraction ranging from a few percent 
--for the few spectroscopically confirmed leakers--
to a few tens of percent --from indirect measurement of the covering fraction of 
low-ionisation state (LIS) absorption lines. 
They certainly produce copious amounts of \lya\ radiation.

A galaxy typically contains several star clusters, 
which means that a LyC leaker may be a mixture of non-leaking and leaking star clusters,
with these two scenarios:
either leaking from a LyC optically thin star-forming region, 
or from an \hii\ region with a low covering fraction.
In the following, we first investigate the \lya\ spectrum 
that emerges from LyC optically thin \hii\ regions 
and the \lya\ spectrum that emerges from riddled LyC optically thick regions, 
both modelled with spherical shells of neutral gas surrounding a \lya\ source.

\subsection{\lya\ transfer in a LyC optically thin  \hii\ region}
\label{s_densitybounded}

We first estimate the optical depth seen by a \lya\ photon 
in a medium that is optically thin for LyC, 
that is, when $\tau_{\rm ion} \lesssim 1$.
The column density of neutral hydrogen along the line 
of sight is of the order of
\begin{equation}
N_{\rm HI} = 1/\sigma_{\nu_0} \sim 1.6\times 10^{17} {\rm cm}^{-2}
,\end{equation}
corresponding to $\tau_{ion} = 1$. 
For comparison, the \lya\ optical depth at the line centre 
of such a low column density medium is still
\begin{equation}
\label{eq_tau0} 
\tau_0 =  5.88 \times 10^{-14} (12.85/b)~N_{\rm HI} \sim 10^{4}
\end{equation}
for a Doppler parameter of $b=12.85$\,\kms, corresponding to an ISM 
temperature of $T = 10^4$K \citep[e.g.][]{Zheng02}.

\subsubsection{Experimental setup}

\begin{figure}
\includegraphics[width = 0.48\textwidth]{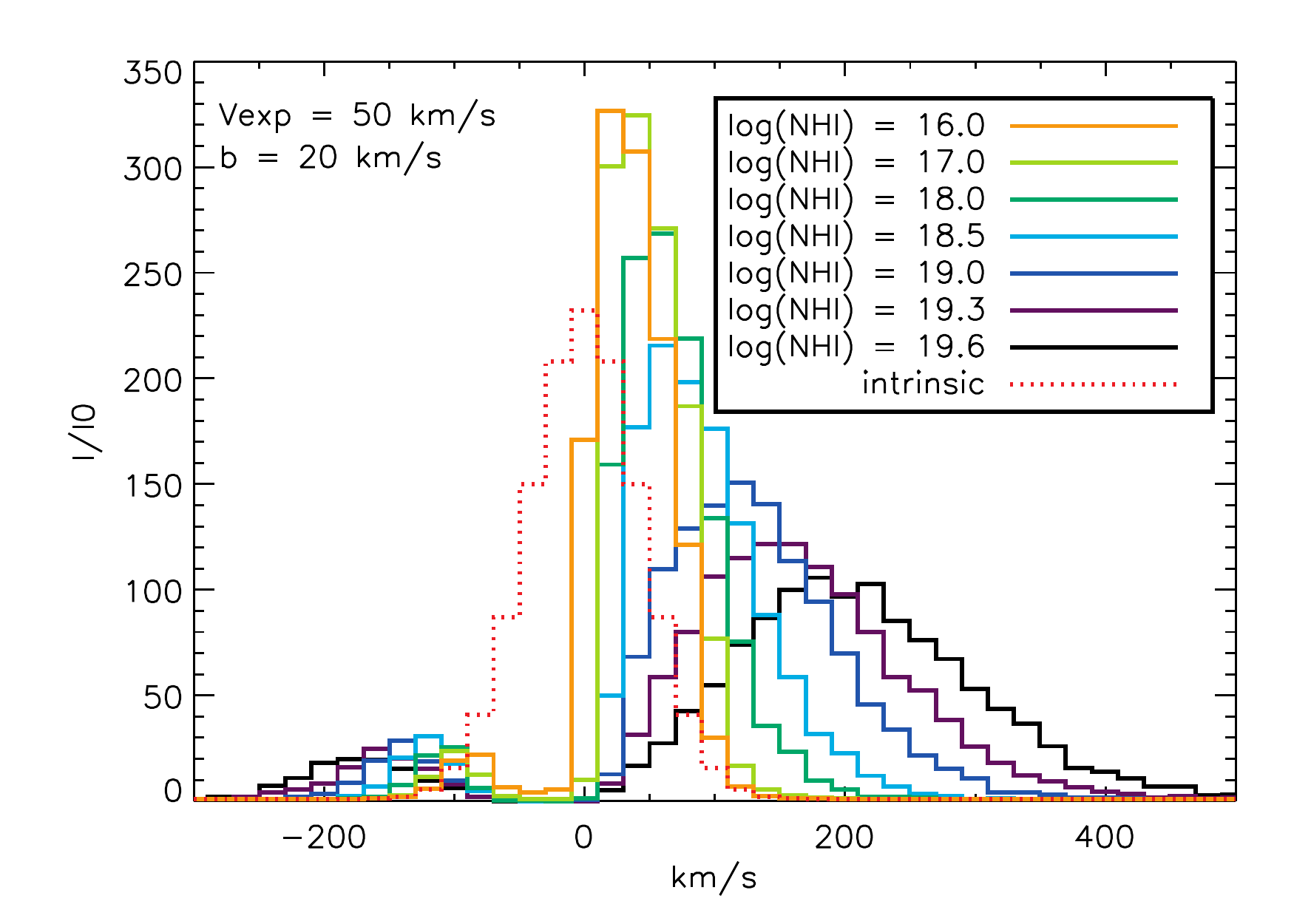} 
\caption{Spectra emerging from starbursts with neutral column densities 
ranging from the LyC optically thin regime (\nh\,$\leq 10^{17}$\,cm$^{-2}$) 
to \nh\,$ = 10^{19.6}$\,cm$^{-2}$, as an illustration of the line broadening with increasing optical depth. 
The models are dust free, with an expansion velocity $\vexp = 50$ \kms, 
and a Doppler parameter $b=20$\,\kms.
The intrinsic spectrum is a flat continuum plus a Gaussian emission line of 
\fwhm\,=\,100\,\kms\ and \ew\,=\,100\,\AA. These spectra are not convolved 
to account for a finite instrumental resolution, but they would
correspond to extremely high-resolution spectra ($R \geq 15000$ or $\Delta v\leq20$\,\kms).
In this plot and in the following theoretical spectra, the notation 
$I/I_0$ indicates that the profiles are normalised to the flux level in the continuum, 
$I_0$, which is the mean number of escaping photons per frequency bin over an arbitrary 
frequency range far from line centre, here xr=[$-3000$\,\kms; $-2000$\,\kms].}
\label{density_bounded}
\end{figure}

The formalism developed by \citet{Neufeld90} is not valid in this 
peculiar regime where $\sqrt{\pi}\tau_0 \leq 10^3/a$, 
with $\tau_0$ the \lya\ optical depth at the line centre and $a$ 
the ratio between the natural to Doppler broadening,
as illustrated for example in Fig.~2 of \citet{Verhamme06}. 
Numerical experiments are necessary to study \lya\ radiation transfer 
in such low optical depth regimes. 
Assuming that \hii\ regions are cavities of ionised gas that  
contain young stars and are surrounded by neutral gas,
we simulated the \lya\ transfer with our Monte Carlo 
code \McLya\ \citep{Verhamme06, Schaerer11} through the same geometrical 
configurations as previously published: spherical shells of homogeneous 
neutral gas that are characterised by four physical parameters, 
\begin{itemize}
\item the radial expansion velocity \vexp,
\item the radial column density \nh, 
\item the Doppler parameter $b$, encoding thermal and turbulent motions,
and\item the dust absorption optical depth $\tau_d$, linked to the extinction 
by \ebv\,$\sim (0.06 ... 0.11) \tau_d$.
\end{itemize}
The lower numerical value corresponds to 
an attenuation law for starbursts according to \citet{Calzetti00}, 
the higher value to the Galactic extinction law from \citet{Seaton79}.
However, the exact geometry that we consider is not important.
We tested that our main results, which we detail below, 
are not altered by other simple geometries such as a sphere or a slab.

\subsubsection{Results: identifiable features in the \lya\ spectrum}

\lya\ radiation transfer in such a low optical depth medium is characterised 
by extremely narrow line-profiles, as illustrated in 
Fig.~\ref{density_bounded}, where we compare emerging \lya\ spectra from 
expanding shells of increasing optical 
depth. The intrinsic spectrum is a Gaussian emission line of \fwhm\,=\,100\,\kms\ and 
\ew\,$=\,100$\,\AA. The shells are homogeneous, isothermal ($b=20$\,\kms),
and dust free, 
and the expansion velocity is 50\,\kms. 
The shift and broadening of the profile is clearly visible when the 
column density increases. 
Extremely narrow \lya\ profiles (\fwhm\,$\sim 200$\,\kms), with a peak shifted by less than 
$\sim 150$\,\kms\ compared to the systemic redshift of an observed galaxy, are therefore 
good candidates for LyC leaking through the LyC optically thin scenario.  

\begin{figure*}[ht]
\begin{center}
\begin{tabular}{ll}
\includegraphics[width = 0.48\textwidth]{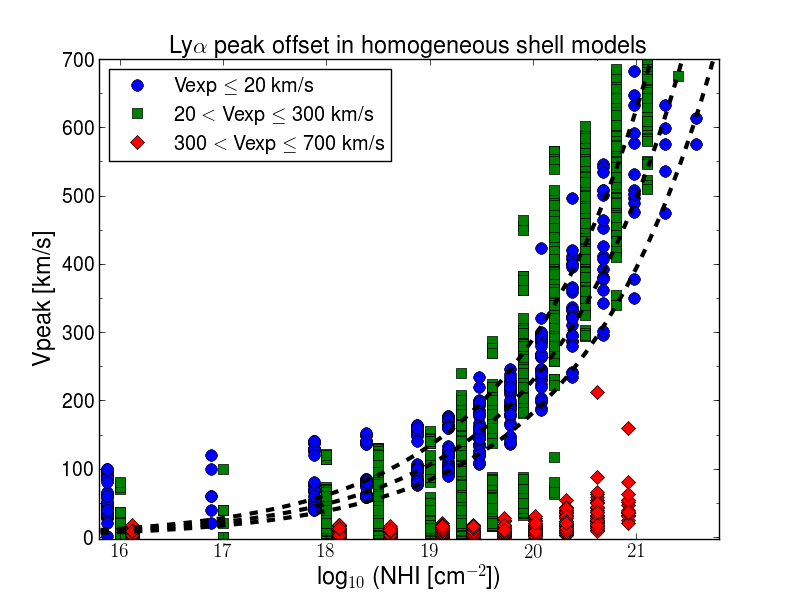} &
\includegraphics[width = 0.48\textwidth]{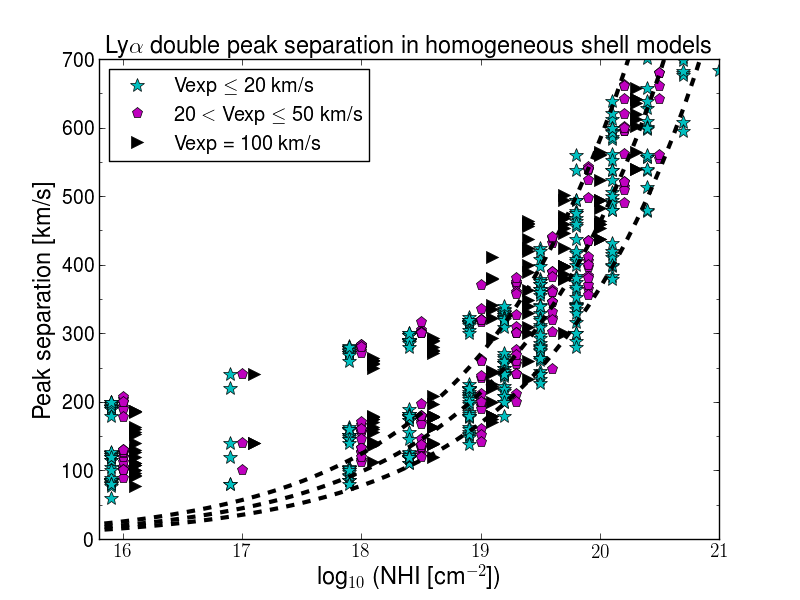} \\
\end{tabular}
\end{center}
\caption{Correlation between the \lya\ peak positions and the neutral gas 
column density as measured in the grid of \McLya\ radiation transfer models 
in spherical homogeneous shells \citep{Verhamme06,Schaerer11}.  
{\bf Left:}  \lya\ peak offset \vpeak. We colour-code three ranges of the outflow 
velocity: static models in blue, moderate outflow velocity models in green, 
high-velocity wind models in red.
{\bf Right:} Separation $S_{\rm peak}$ of peaks in the double-peak \lya\ profiles, i.e.
those resulting from \lya\ radiative transfer in low-velocity outflows 
 (the profiles become single-peaked for velocities higher than 100\,\kms).
The black curves in both panels depict the analytically derived \lya\ peak positions for 
radiative transfer in a static uniform sphere \citep{Dijkstra06}.
From top to bottom, they correspond to $b=40$\,\kms, $b=20$\,\kms\ and $b=10$\,\kms.
Some discrepancy between the analytical estimate of Vpeak and the numerical experiment is 
expected at the low NHI end, since the analytical solution is only valid in an optically 
thick regime \citep[see for example Fig.2 in ][]{Verhamme06}.       
We checked that these results are independent of instrumental resolution. 
}
\label{vpeak}
\end{figure*}

In the left panel of Fig~\ref{vpeak}, the offset between the profile maximum 
and the centre of the restframe \lya\ line versus the neutral 
column density for a sample of shell models extracted from the library 
described in \citet{Schaerer11}, with moderate Doppler parameter ($10 < b < 40$ \kms),
and moderate extinction ($0. < \ebv < 0.2$), to restrain our study to 
\lya\ profiles in emission.
The global trend is a correlation between log(\nh) and \vpeak: 
when the column density of the shells increases, the peak of the profile 
is increasingly redshifted away from the line centre, except for high values 
of the expansion velocity (\vexp$ > 300$ \kms, red dots in the plot). 
The reason why models with high outflow velocity have a small shift 
of the \lya\ peak is that a \lya\ photon arriving at the fast moving shell 
will be seen out of resonance, and its probability to cross the shell 
without interaction is high. 
For high enough velocities, the intrinsic, unaltered \lya\ spectrum would be recovered
by the observer.

\begin{figure*}
\begin{center}
\begin{tabular}{cc}
\includegraphics[width = 0.48\textwidth]{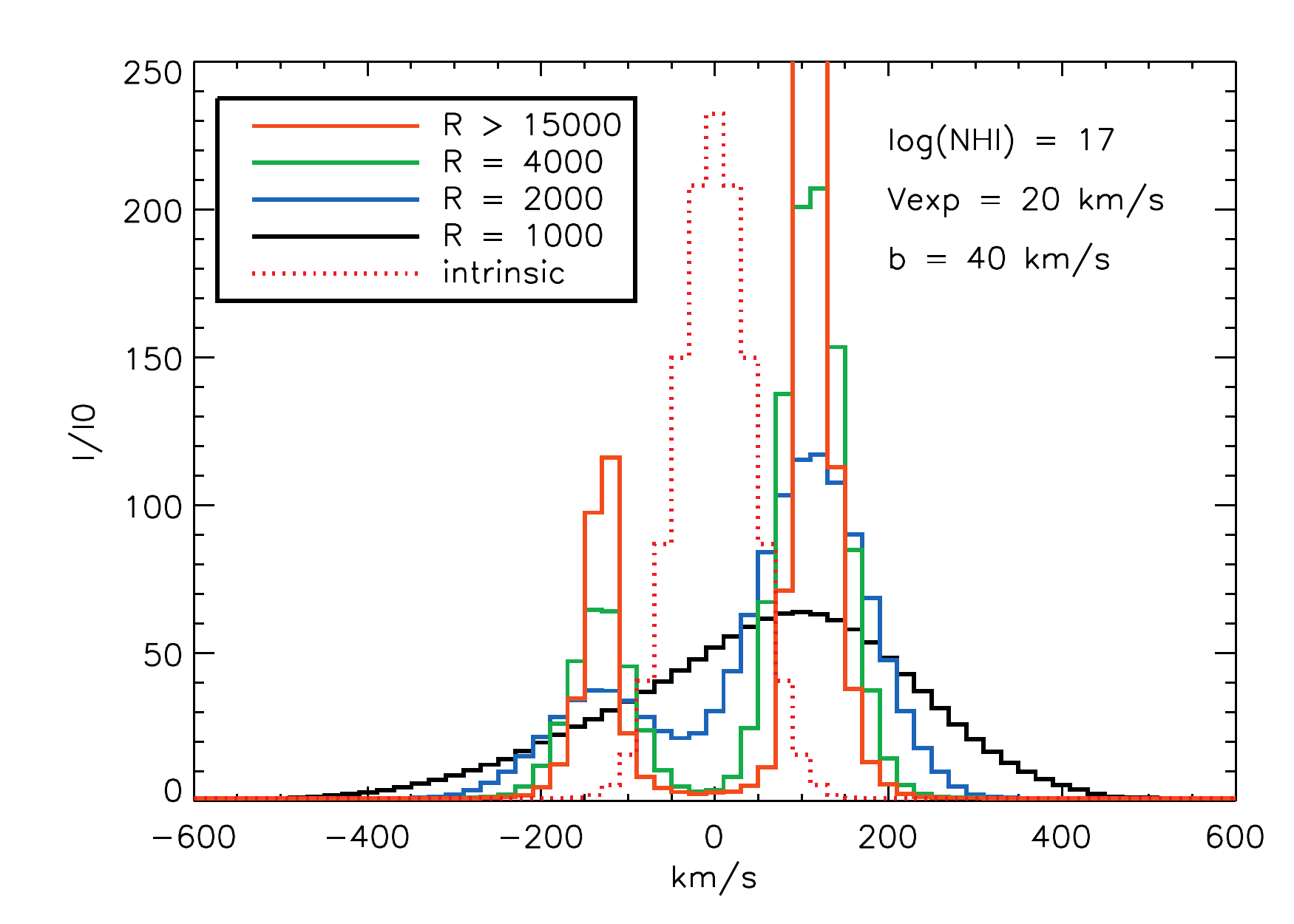} &
\includegraphics[width = 0.48\textwidth]{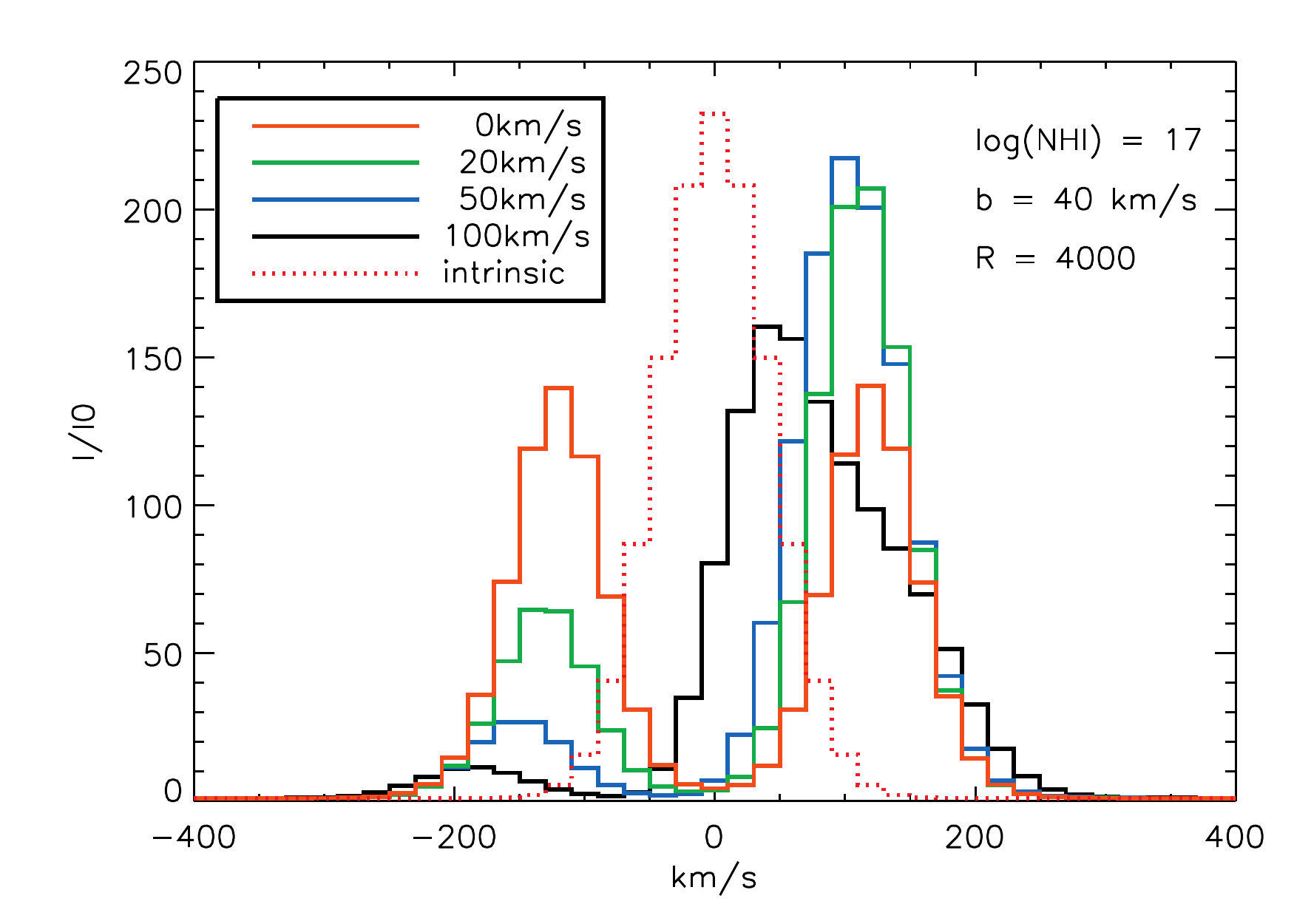} \\
\end{tabular}
\end{center}
\caption{
Influence of the spectral resolution (left) and the outflow velocity 
(right) on the shape of emerging spectra from density-bounded starbursts with 
low neutral column densities (N$_{\rm HI} \leq 10^{17}$cm$^{-2}$).
An observational signature of density-bounded 
ISM is a small shift of the profile maximum, \vpeak, 
independent of the expansion velocity \vexp.}
\label{lowdens_Resol+vexp}
\end{figure*}

Figure\,\ref{vpeak} clearly shows that models with \nh\,$ < 10^{18}$\,cm$^{-2}$ are
characterised by \vpeak\,$<150$\,\kms. However, it also shows that the low \vpeak\
is not a sufficient condition for identifying a LyC leaker, because other effects 
(high outflow velocities) may cause low \vpeak, too.
Nevertheless, \lya\ spectra are very useful for\textup{ {\em excluding}} LyC leakage:
a large peak offset, \vpeak\,$>150$\,\kms, indicates that \hi\ column densities are 
too high to allow LyC escape.
In contrast, \lya\ spectra with a low \vpeak\ identify {\em \textup{candidates}} for LyC leakage
in optically thin ISM, which
need to be confirmed by independent indicators or by direct LyC observation.
An estimate of outflow velocities from the blueshift of metallic low-ionisation
state (LIS) absorption lines should also be able to distinguish between objects
with low \vpeak\ due to high \vexp, and objects with a low
\vpeak\  due to a low column density.
We verified that the location of \vpeak\ is not very sensitive 
to the spectral resolution (see the left panel of Fig.~\ref{lowdens_Resol+vexp}).
The width of the \lya\ peaks would be yet another characteristic
that is dependent on the \hi\ column density. However, the line width is also 
affected by spectral resolution, and therefore we provide no selection 
criteria  for this parameter.

\bigskip

The \vpeak\ reliability critically depends on the redshift precision. 
We therefore propose a complementary criterion, independent of redshift: 
in double-peaked \lya\ profiles, a small peak separation ($S_{\rm peak} < 300$\,\kms)
is a good tentative indication of transfer in a medium with a low column density, 
as illustrated in the right panel of Fig.~\ref{vpeak}. 
Similar to the \vpeak\ prediction, \speak\ increases with \nh, and 
a peak separation $>300$\,\kms\ rules out a low optical depth of the ionising radiation.  
However, \speak\ $<300$ can exist for \nh\ as high as 
$10^{20}$ cm$^{-2}$. A small peak separation is a good indication of a low column density,
but not a definitive probe of LyC leaking.
Although the method is simple, it holds only for double-peaked spectra, 
which appear in static to slowly expanding media: $\vexp \leq 100$\,\kms\
from our library. 
Synthetic spectra with higher \vexp\ have a weak blue peak, 
which diminishes as \vexp\ increases
(see the right panel in Fig~\ref{lowdens_Resol+vexp}).

We identify a potential difficulty when using this prediction
on observed spectra with unknown redshift: the peaks are supposed to be 
at each side of the line centre, which is not always the case according to the sample of \citet{Kulas12}.
Furthermore, some observed spectra have complex substructures 
(bumps or multiple peaks, see Sect.~\ref{s_compare}), 
and the main peak location may be difficult  to determine.

\subsubsection{Discussion}

High spectral resolution is needed to recover extremely narrow profiles. 
The actual canonical good spectral resolution at high redshift is typically 
$R \sim 3000$ on VLT-FORS or the Mitchell spectrograph \citep{Chonis13, Tapken07}, 
or even $R \sim 5000$ with Xshooter on lensed targets \citep{Christensen12a,
Christensen12b, Noterdaeme12}. 
In the left panel of Fig.~\ref{lowdens_Resol+vexp}, we illustrate the broadening 
of a theoretical \lya\ profile that is due to instrumental resolution. 
The orange sharp profile is the theoretical one, unconvolved, labelled $R>15000$
since the binning corresponds to $\Delta v\leq20$\,\kms. 
The green curve has been convolved with a Gaussian of \fwhm\,=\,75\,\kms\ corresponding 
to an observation of the orange profile with a spectral resolution of $R=4000$.
The peaks are slightly broader, but the shape of the profile is relatively well conserved. 
The blue profile corresponds to $R=2000$.
The two peaks are still identifiable, but much broader; 
the central absorption disappears, the whole profile is detected in emission.
The black curve corresponds to $R=1000$.
The peaks are not identifiable anymore, and the spectrum presents a blue bump,
or is elongated towards the blue.
When the spectral resolution decreases, the location of the main peak of 
the profile, \vpeak, is rather well conserved. 
It slightly shifts towards lower values.
\vpeak\ is much less sensitive to the spectral resolution than the peak width
or the overall spectral shape. 

By comparing dust-free and dusty models of 
density-bounded ISMs, we checked that dust only has a weak effect 
on the \lya\ profiles: the static case is only affected by a 
global flux reduction, but the double-peaked shape is conserved. 
When the medium is non-static, the effective \lya\ optical depth 
decreases strongly and the transfer becomes almost insensitive to the 
dust content. The location of \vpeak\ does not depend on $\tau_d$.

\bigskip

Neutral column densities as low as $10^{17}$cm$^{-2}$ are certainly unusual 
for the intervening interstellar medium of galaxies. 
Peculiar lines of sight, aligned by chance with an outflow axis, or peculiar 
objects, with extremely high ionisation parameters \citep[e.g. blue compact galaxies, 
green peas, extremely strong emission line galaxies;][]{Cardamone09,Izotov11,Jaskot13}, 
may present these characteristics, however.
Observationally, this high-ionisation parameter can be estimated by 
measuring the \oiii/\oii\ ratio \citep{Nakajima14, Jaskot14}.
A high value of \oiii/\oii\ ($>10$) can result from several factors 
\citep{Kewley13, Stasinska15}: 1) density-bounded H{\sc ii} regions with 
reduced outermost [O {\sc ii}] regions; 2) a high-ionisation parameter (the
high-ionising flux enhances \oiii\ relative to \oii); and 3) a low metallicity.
In cases of \lya\ transfer through density-bounded regions, we expect both
\vpeak $\leq 150$ \kms\ and a high \oiii/\oii\ ratio.

Another observational piece of evidence for a low column density would be to consider the 
metallic low-ionisation state absorption lines \citep[e.g.][]{Heckman11, Jones13}. 
Saturated lines imply a high covering fraction of neutral gas along the line of sight. 
In contrast, weak or undetected LIS absorption lines may be a good indication for 
a low column density \citep{Erb10, Jaskot14}.
However, this indicator is very sensitive to spectral resolution
as well as metallicity.

\subsection{\lya\ transfer in a riddled ISM} 
\label{s_riddled}

Another scenario to allow Lyman-continuum radiation to escape the ISM of a
normal galaxy (N$_{\rm HI} \gg 10^{17}$cm$^{-2}$) would be a partial coverage 
of star-forming regions with neutral gas. 
This would also have consequences on the emerging \lya\ profiles.

\subsubsection{Experimental setup}

Detailed studies of \lya\ radiation transfer in clumpy media have recently been conducted 
\citep{Dijkstra12, Laursen13, Duval14, Gronke14}. 
They focused on regimes in which the resonant \lya\ radiation will escape better than the stellar UV continuum, 
which suffers from extinction. The main result from these studies is 
that \lya\ can escape from extremely optically thick and dusty media 
in very particular cases: a rather static ISM, with two phases strongly 
contrasted in density, and a very high covering fraction of the dense clumps.
These studies can be seen as numerical validations of the predictions from 
\citet{Neufeld91}. 

We here consider the same kind of two-phase media, static or not,
but with a different focus: we search for Lyman-continuum-leaking regimes,
or in other words, for clumpy media with a non-unity covering fration of the neutral gas 
(CF $\leq 98$\%), and we observe them along LyC-transparent lines of sight.
We here adopted the clumpy medium and spherical geometry described in \citet{Duval14}:
the shells were built on a $128^3$ Cartesian grid, with a minimum 
(maximum) radius of Rmin = 49 (Rmax = 64), and a filling factor FF = 0.23, 
set to best fit the observed mass spectrum of the ISM clumps \citep{Witt00}. 
For simplicity, we considered here that the density 
of the interclump medium is zero   \citep[“high contrast regime” in][]{Duval14}, 
to allow for LyC escape through transparent lines of sight even if in principle 
the interclump medium could have a residual neutral column density 
N$_{\rm HI} \leq 10^{17}$cm$^{-2}$. These clumpy models are illustrated in Figs.~1 
and 2 of \citet{Duval14}. As an example, the mean number of clumps along a sight 
line is $\sim 4$ in a medium with $CF = 0.90$.

\subsubsection{Results: identifiable features in the \lya\ spectrum}

\begin{figure*}
\begin{tabular}{cc}
\includegraphics[width = 0.48\textwidth]{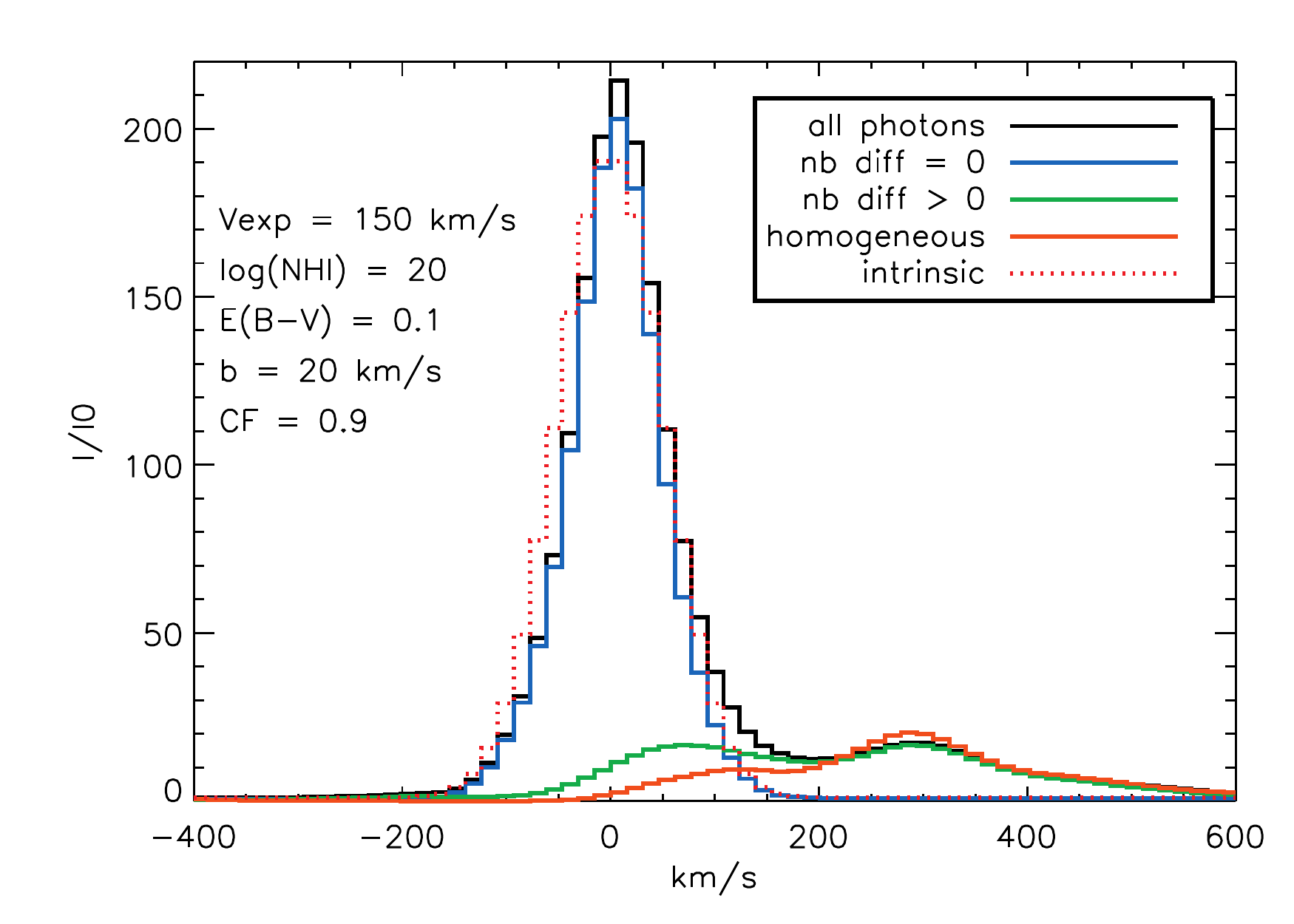} &
\includegraphics[width = 0.48\textwidth]{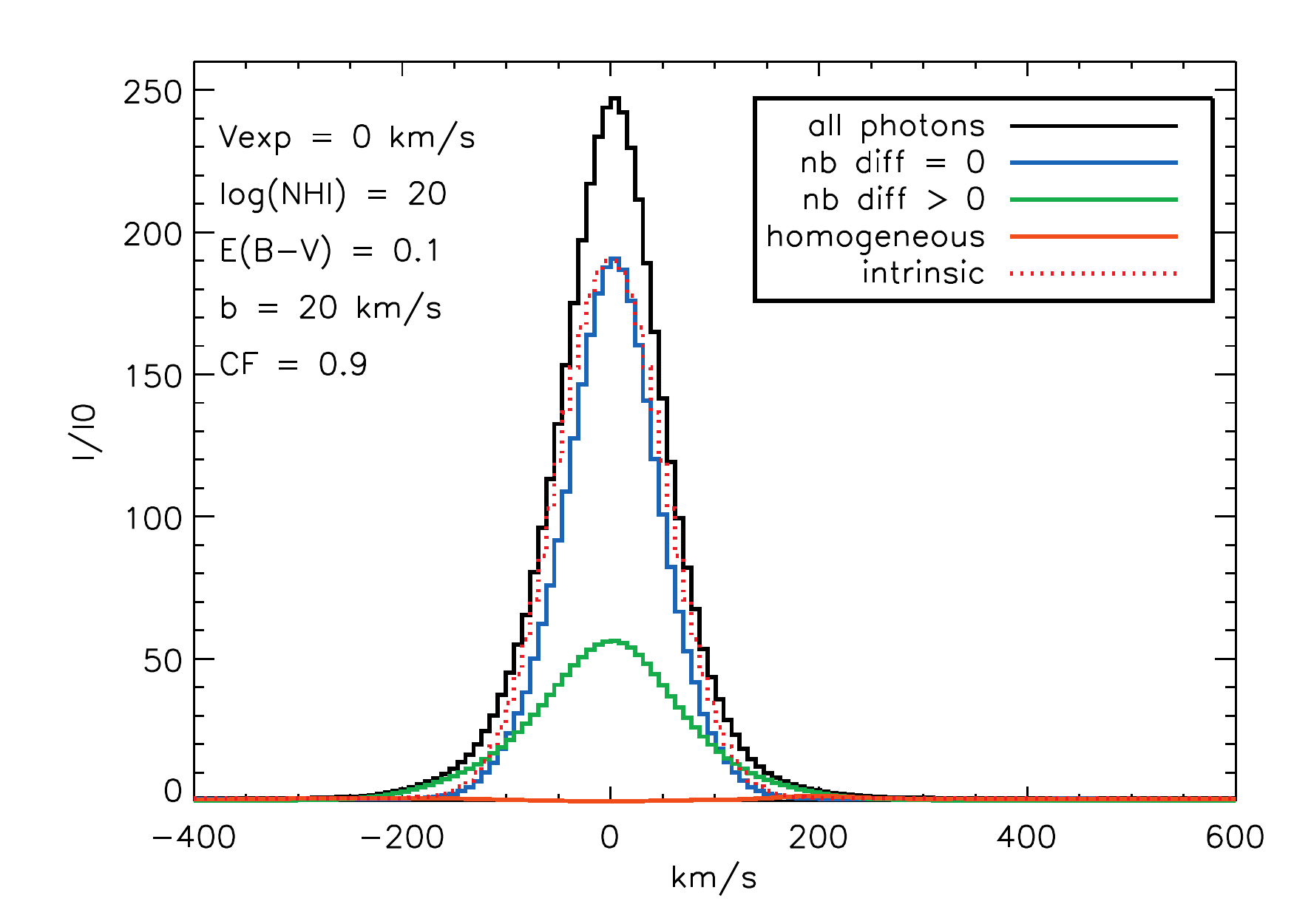} \\ 
\end{tabular}
\caption{
 \lya\ spectra emerging from clumpy spherical shells with non-unity 
coverage. 
{\bf Left:} Expanding shell with fiducial values of the parameters 
(\vexp\ = 150\,\kms, log(N$_{\rm HI}) = 20$, b = 20\,\kms, \ebv\,=\,0.1). 
{\bf Right:} Static shell with exactly the same values of the parameters, 
except the velocity field, which is set to zero.
In both cases, the intrinsic spectrum is a continuum plus a Gaussian emission 
line of \fwhm\,$= 100$\,\kms\ and \ew\,$= 100$\,\AA\  (red dotted line). 
The black lines shows the global emerging spectra, spatially integrated over the 
whole shell. The blue and green lines are a decomposition of the black curves 
in photons that escaped without scattering ($nb_{\rm diff} = 0$, blue) 
and photons that interacted before escape ($nb_{\rm diff} > 0$, green). 
 This second packet of photons is compared to the result of \lya\ 
 radiation transfer in a homogeneous shell (orange curve).
These spectra are convolved with a Gaussian to mimic an instrumental resolution of $R=4000$.
The main features in the \lya\ spectra that emerge from transparent lines of sight in riddled 
ionisation-bounded \hii\ regions are \vpeak\ = 0 and, as a consequence, a non-zero flux 
bluewards of the systemic redshift.
}
\label{ionisation_bounded}
\end{figure*}

In clumpy geometries, the spherical symmetry is broken: all lines of sight are not 
equivalent in terms of LyC escape. Peculiar lines of sight, where the source is aligned 
with a hole, will emit ionising flux, whereas other directions, where the source is 
obscured, will not emit any ionising photon. This study focuses on the link between 
\lya\ and LyC escape. In the following, we then consider only \lya\ spectra that emerge 
from LyC transparent lines of sight, that is, with a hole in front of the source.

In Fig.~\ref{ionisation_bounded}, we present the \lya\ spectra emerging from 
clumpy spherical shells with a spatial covering fraction, CF, below unity, 
observed along a transparent line of sight along which ionising photons can escape. 
The spectra correspond to high-resolution spectra (R=4000).
The intrinsic spectrum is a continuum plus a Gaussian emission line of 
\fwhm\,$= 100$\,\kms\ and \ew\,$= 100$\,\AA.
In the left panel we show the emerging profile from a clumpy shell 
with the following fiducial values for the four physical parameters:
\vexp\ = 150\,\kms, log(N$_{\rm HI}$) = 20, $b = 20$\,\kms, and \ebv\,=\,0.1
These values correspond to the most common best-fit values in our studies 
fitting observed spectra with our library of homogeneous shells 
\citep[][Orlitov\'a et al. in prep; Hashimoto et al. in prep]
{Verhamme08, Dessauges10, Lidman12}. 
We chose CF=0.9 as a fiducial value for the covering factor because it translates
into an escape fraction of the ionising continuum of 10\%, which corresponds to a high
value among the few measurements of LyC escape from galaxies.  
In the right panel we show  \lya\ spectra that emerge from a static shell with the same 
fiducial values except for the expansion velocity, which is set to zero,
as well as the profile from a homogeneous shell (orange curve).

As is clear from Fig.~\ref{ionisation_bounded}, the \lya\ profiles from transparent 
sight-lines in a clumpy, spherical shell with a spatial coverage below unity show most 
of the flux emerging at the line centre (black curve). \citet{Behrens14} also found the 
same result: their spectra escaping along lines of sight with holes look very similar 
to ours, with a very prominent, centred emission (e.g. their Fig. 7).
We divided the emerging spectrum from a clumpy medium into two components: 
{\em 1)} the spectrum of photons that did not undergo any scattering before escape 
($nb_{\rm diff} = 0$, blue curve), which is identical to the intrinsic spectrum 
(red dotted line), independently of the geometry or the velocity of the scattering medium, 
{\em 2)} a second component, made of all photons that scattered before escape 
($nb_{\rm diff} > 0$, green curve), which will then depend on the geometry and velocity 
of the scattering medium. However, this component has a different shape than the spectrum 
emerging from a shell with the same fiducial values of the parameter, but homogeneous 
(orange curve). 
Especially in the static case, the spectrum that emerges from a homogeneous shell shows a broad 
absorption Voigt profile plus some residual double-peaked emission (orange curve,
 although the scale is adjusted to see the strong emission peak in other curves, 
and the absorption feature is barely visible), whereas the spectrum that emerges from the 
clumpy shell after scattering shows a symmetrical and centred emission line (green curve). 
We interpret this difference between the green and orange curves seen in both experiments, 
but enhanced in the static case, as due to \lya\ radiation tranfer effects: in a medium 
with holes, resonant photons can bounce on the clumps and find the holes to escape, 
whereas in the homogeneous medium, they strongly diffuse and increase their chance of being 
absorbed by dust.
Note that the spectrum that emerges along a line of sight without a hole in front of the source
 will look like the green curve, meaning that it will be only composed of photons having scattered 
through this clumpy medium. 

As a consequence of adding this diffuse component (green curve) to the 
intrinsic profile, the equivalent width of the spectrum emerging
along a LyC tranparent 
line of sight is then always higher than the intrinsic \lya\ equivalent width.
We therefore predict that \lya\ spectra with a peak at the systemic redshift and, as a 
consequence, a strong \lya\ emission and also a non-zero flux bluewards of the \lya\ line 
centre, are good candidates for LyC-leaking objects that leak
as a result of a porous ISM.
Obviously, partial coverage of neutral gas along the line of sight
can also be detected by measuring low-ionisation state (LIS)
metal absorption lines \citep[e.g.][]{Jones13}.

\begin{figure}
\includegraphics[width = 0.5\textwidth]{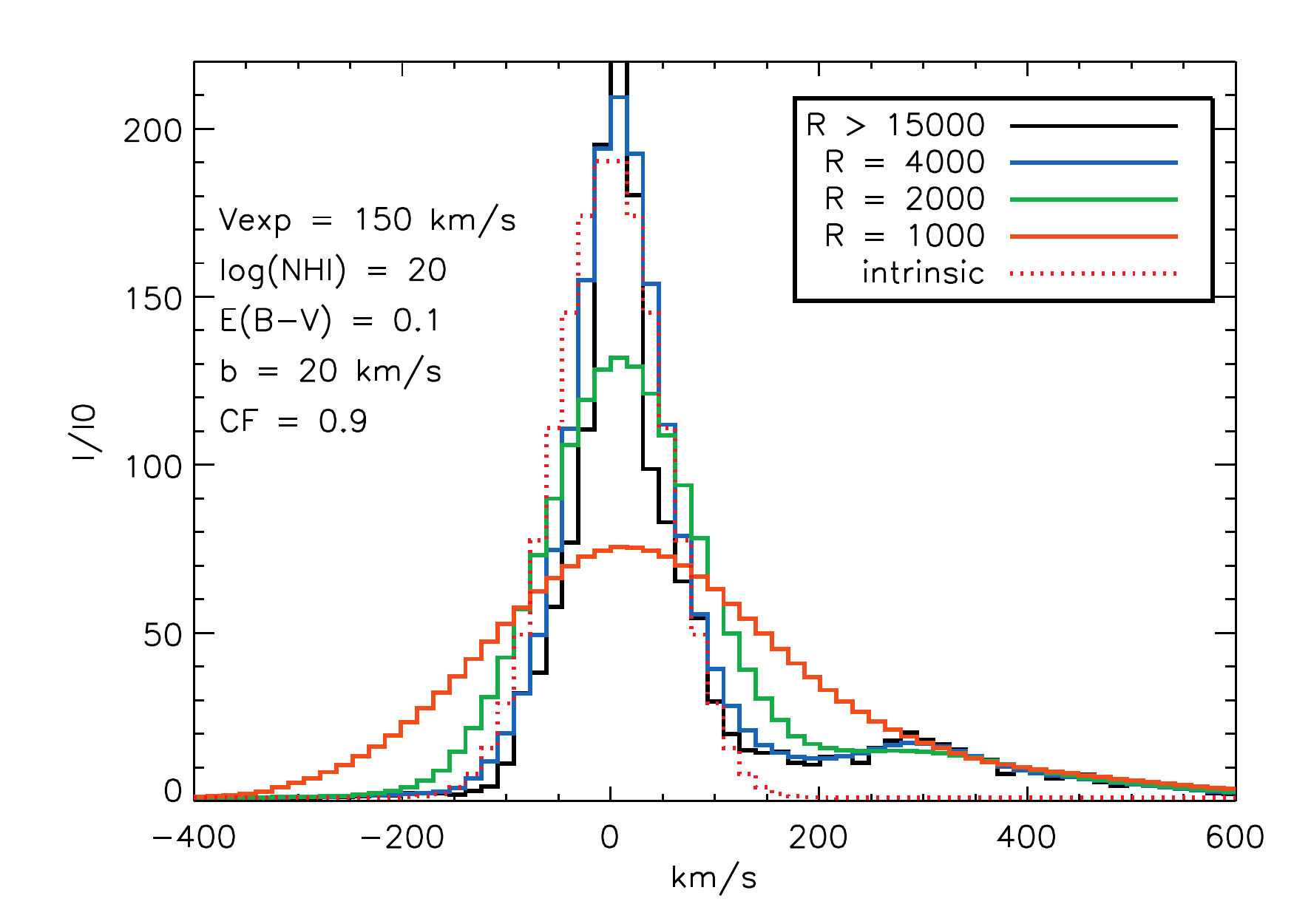} 
\caption{Effects of spectral resolution on the \lya\ spectra that emerges from a 
clumpy shell with fiducial values for the four physical parameters, 
and a covering fraction of 90\%.}
\label{clumpy_resol+vexp}
\end{figure}

In Fig.~\ref{clumpy_resol+vexp}, we present the effect of spectral
resolution on the two diagnostics that we propose (peak at $v = 0$\,\kms\ and
non-zero flux bluewards of \lya) to identify LyC leakers. 
Although a degradation of the spectral resolution leads to 
an overestimate of the true emerging flux bluewards of \lya\
and \vpeak\ slightly shifts towards the red,
the profiles peak at or close to  $v = 0$ 
and show a non-zero flux bluewards of $v = 0$,
quite independently of \vexp.
Hence, our diagnostics for a clumpy Lyman-continuum-leaking ISM
are largely independent of the outflow velocity
and  can be detected with present spectroscopic observations
as long as the covering fraction is below unity.

\subsection{Summary of the link between \lya\  and LyC}

\subsubsection{\lya\ spectral diagnostic for LyC leakage}

We have discussed two cases of LyC leakage: 
from \hii\ regions with low column densities of neutral gas,
and from \hii\ regions with partial coverage of neutral gas.
For the first scenario of LyC leakage from density-bounded \hii\ regions, 
we proposed that observed \lya\ spectra with a typical asymmetric redshifted profile 
are good LyC leaking candidates if the peak of the profile, \vpeak, is within $150$\,\kms
of the systemic redshift.
We have shown that this diagnostic is not very sensitive to the spectral resolution.
When there is a blue peak in the profile, a peak separation smaller than 
$\sim300 (2 \times 150)$\,\kms\  may also be a good indicator for \lya\ transfer in media with a low 
column density (log(N$_{\rm HI}$) $\leq 18$). 

For the second scenario of LyC leakage, we claimed that \lya\ spectra with 
a main peak at the systemic redshift of the object, and non-zero flux 
bluewards of the line centre, are good candidates for continuum leaking. 
Despite the differences in the geometries studied for instance
in \citet{Behrens14} and 
here, that is, anisotropic vs. isotropic distributions of gas with a non-unity covering fraction, 
we reached the same conclusion: lines of sight containing holes lead to \lya\ profiles 
with a prominent peak at the line centre.

\subsubsection{Limitations of the method}
Not all LyC leakers are expected to show this type of \lya\ spectra.
In particular, the very efficient leakers, with a high escape fraction 
($f_{\rm esc}(LyC) \geq 80$\%), may have a low \lya\ \ew, or even no \lya\
emission at all when $f_{\rm esc} = 1$ \citep[see Fig.~13 in][]{Nakajima14}, 
but they may be rare \citep{Vanzella10}.
If most LyC leakers have a moderate $f_{\rm esc}(LyC) \sim 10\%$, 
we then predict that their \lya\ spectra will be peculiar, as described above.

We did not take into account the effect of IGM attenuation in our modelling. 
The neutral fraction of the intergalactic medium increases with redshift. 
The remaining neutral gas in the circumgalactic/intergalactic medium around
a galaxy will scatter the light bluewards of \lya\ out of the line of sight.
The IGM transmission has been predicted to reach a minimum near 
line-centre \citep{Laursen11}, which may alter the line shape by artificially 
redshifting the peak.  This effect may complicate the usage of this 
method at high redshift. However, if they are detected, high-redshift \lya\ 
spectra with small \vpeak\ will be the best candidates for continuum leaking.

Other configurations or physical processes, such as clumpy geometries 
with CF=1 in the peculiar Neufeld scenario, high outflow velocities, 
fluorescence, or gravitational cooling could lead to the same \lya\ 
profiles without leaking.
\citet{Laursen13} and \citet{Duval14} discussed in detail the conditions of appearance of 
an enhancement of \lya\ escape compared to the UV (non-ionising) continuum. 
They argued that the peculiar conditions for boosting \lya\ compared to continuum 
appear improbable in the interstellar medium of galaxies. 
However, the resulting \lya\ profile shown in Fig.~17 in \citet{Duval14} is a narrow line, 
perfectly symmetric and centred on $v = 0$, similar to our prediction of LyC leaking 
through the second scenario. But in this case, only \lya\ can escape, not LyC, 
since the covering fraction is unity. 
A way to distinguish between our scenario of LyC leakage
and the unlikely Neufeld scenario would be to consider the LIS absorption lines, 
which should be saturated, and not blueshifted but centred on the systemic redshift, 
tracing a static medium with CF=1.

If the expansion velocity of the neutral gas around a starburst 
reaches high values \citep[$\vexp > 300$\,\kms\ , as observed in][]{Bradshaw13, Karman14}, 
\lya\ photons may escape through the neutral shell without scattering,
 and the main peak of the \lya\ profile may be located at or close to the systemic redshift 
(red points in Fig.~\ref{vpeak}), but no ionising photons will be able to escape. 
The precise \vexp\ beyond which the shell becomes transparent ($\tau_0 \sim 1$) 
can be expressed as a function of $N_{\rm HI}$:
\begin{equation}
\vexp = \sqrt{\frac{4.7 \times 10^{-17} N_{\rm HI}}{\pi}} \, T_4^{-1/2} \, b. 
\end{equation}
This equation is valid when $\vexp/b = x_c > 3,$ and the Hjerting function can be 
approximated by $H(x,a) \sim \frac{a}{\sqrt{\pi} x^2}$. 
An estimation of this outflow velocity from the blueshift of LIS 
absorption lines should allow distinguishing between high-velocity or 
low column density cases.

\citet{Cantalupo05} and \citet{Kollmeier10} predicted the \lya\ spectra 
and luminosities emitted by a cloud of neutral gas, enlighted by a close-by quasar, 
or embedded in the cosmological UV background. The external layers of the gas are ionised 
by the external source, recombine, and emit \lya. This process is called \lya\ 
fluorescence. \citet{Cantalupo12} showed convincing detections of LAEs that are powered 
by fluorescence.
Some of the \lya\ spectra observed by \citet{Trainor13} from objects illuminated by 
a strong quasar show a narrow \lya\ profile with a blue bump and a small separation 
between the peaks (Trainor, private communication).
Indeed, in case of fluorescence, the external layer of a cloud of neutral gas is ionised.
The optical thickness of this layer will be $\sim 1$ for ionising radiation, corresponding 
to a neutral column density of N$_{\rm HI} \sim 10^{17}$ cm$^{-2}$. 
The \lya\ photons created by recombination inside this layer will then diffuse 
in such an optically thin medium and should resemble the spectra described 
above in our first scenario of LyC leakage.
In cases of fluorescence, the external source of ionisation is hopefully either known
or bright and hence easily identifiable.

In addition to \lya\ emission from gas recombination around starbursts, 
the other \lya\ production channel is via collisionnal excitation of gas 
falling onto a halo and converting its gravitational energy to heat 
through collisions. \lya\ emission is the main process for this 
almost pristine gas to cool.
However, the \lya\ spectra observed from \lya\ nebulae
attributed to gravitational cooling are usually extremely broad (\fwhm\,$\sim 1000$\,\kms), 
which differentiates them from LyC-leaking candidates, as well as images, 
with extended \lya\ emission over several tens of kpc. 
\citet{Faucher10} computed the \lya\ spectra of cold streams, 
which they find double-peaked, almost symmetrical (note that the IGM was not included 
in these calculations), and broad.


\section{Comparison with other leaking indicators}
\label{s_compare}

\begin{figure}
\includegraphics[width = 0.5\textwidth]{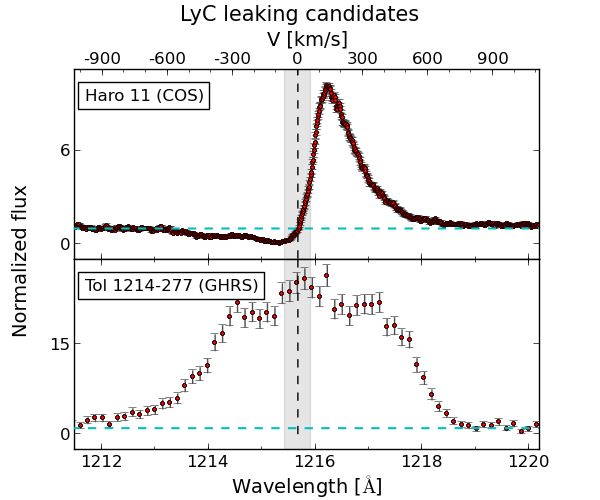} 
\caption{Observed \lya\ spectra of LyC leakers from our \lya\ criteria.
{\bf Top:} HST/COS \lya\ profile of the known LyC leaking galaxy Haro\,11 
\citet{Bergvall06}, with a narrow P Cygni profile with $\vpeak \leq 150$\,\kms, 
typical of \lya\ transfer through low optical depth medium.
{\bf Bottom:} GHRS \lya\ spectrum of a low-metallicity blue compact dwarf 
galaxy Tol\,1214$-$277, 
highly suggestive of Lyman-continuum leakage from a riddled ISM
(Sect.\ \ref{s_riddled}).
The systemic redshifts and their errors
(grey shaded area in each plot) were adopted from the 6dF survey for 
Haro\,11 and from \citet{Pustilnik07} for Tol\,1214$-$277. 
The blue horizontal line denotes the continuum level. }
\label{lya_leakers}
\end{figure}

\begin{figure*}
\begin{tabular}{cc}
\includegraphics[width = 0.5\textwidth]{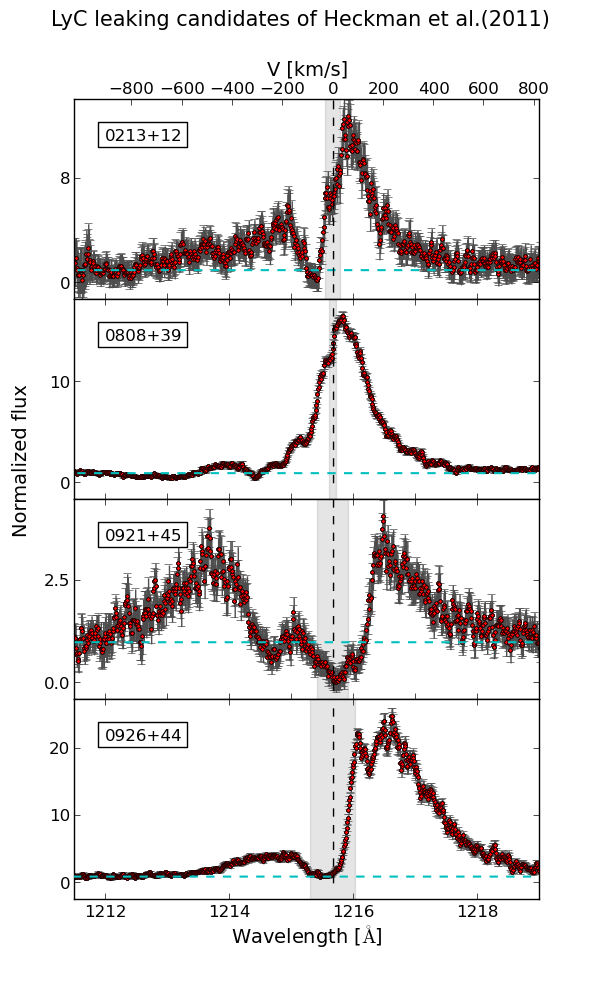} &
\includegraphics[width = 0.5\textwidth]{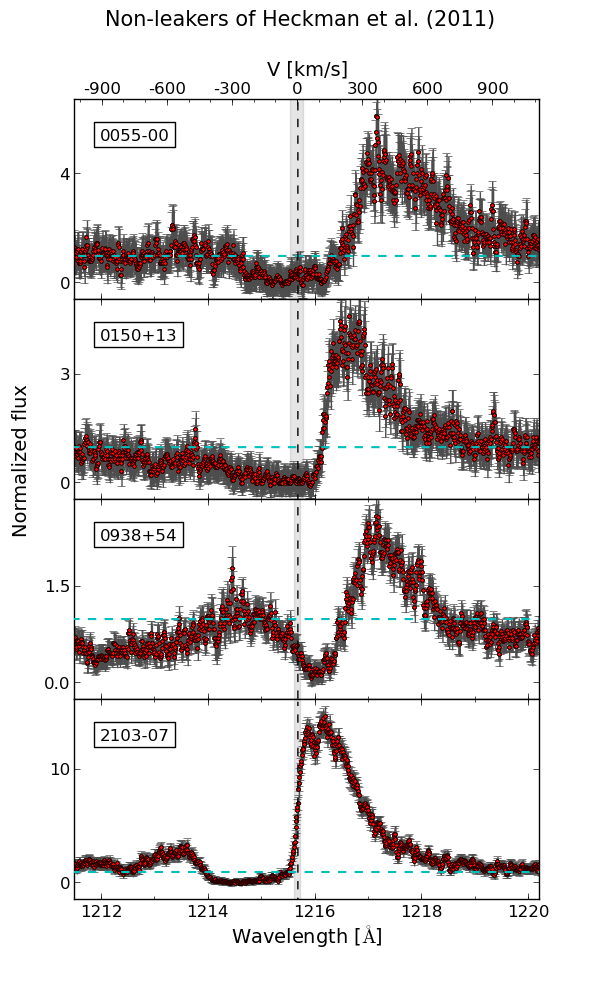} \\ 
\end{tabular}
\caption{\lya\ profiles of LyC leaking (left) vs non-leaking candidates (right) from the 
 sample of \citet{Heckman11}. 
The data have been smoothed to the nominal COS resolution (i.e. $R=20\,000$ for a 
point source). The systemic redshifts and their errors 
(grey shaded area in each plot) were derived from SDSS emission lines.
The blue horizontal line denotes the continuum level. 
The signal-to-noise differences across the sample reflect both the
redshifts and the intrinsic properties of the galaxies as \lya\ sources.
}
\label{CF_leakers}
\end{figure*}

\begin{table*}
\caption{Compilation of several indicators for LyC leakage from the 
low-redshift sample of candidate and confirmed LyC leakers\tablefoottext{1}.}
\label{Tab}
\begin{tabular}{lccccc}
  \hline
  \hline
  object ID\tablefootmark{$\ast$}  & \fesclyc & \fesclyc & \oiii/\oii\tablefootmark{c} & \vpeak(\lya)  & \lya\ peak separation \\ 
                  & from direct LyC detection\tablefootmark{a} & from UV covering fraction\tablefootmark{b} & & [\kms]\tablefootmark{d} & [\kms]\\
  \hline                                                  
  \hline                                                  
  \object{Tol 1247$-$232} & 0.024  &  --    &  3.4 $\pm$ 0.6     & 130 $\pm$ 70   & 500 $\pm$ 30 \\ 
  \object{Haro 11}         & 0.032  &  0.024\tablefootmark{a} &  1.5 $\pm$ 0.2     & 110 $\pm$ 60   & -- \\
  \object{Tol 1214$-$277} &  $>0$ suspected\tablefootmark{e} & $>0$ suspected\tablefootmark{f} & 15 $\pm$ 3 &  60 $\pm$ 60  & 650 $\pm$ 50\\
  {GP 1219+15}      &  $>0$ suspected\tablefootmark{e} &  --  &  10.2 $\pm$ 0.3 & 170 $\pm$ 20 & 
 270 $\pm$ 20 \\
\hline
  {0213+12}         &   --  &  0.05  &  0.8 $\pm$ 0.4    &  70 $\pm$ 30  &  230 $\pm$ 10 \\
  {0808+39}         &   --  &  0.12  &  0.60 $\pm$ 0.08  &  30 $\pm$ 20  &  450 $\pm$ 30 \\
  {0921+45}         &   --  &  0.04  &  0.28 $\pm$ 0.03  & 240 $\pm$ 60  &  690 $\pm$ 10 \\
  {0926+44}         &   --  &  0.40  &  3.2 $\pm$ 0.1    & 280 $\pm$ 90  &  450 $\pm$ 50 \\
  \hline                                                 
  {0055$-$00}       &   --  &$< 0.01$&  1.52 $\pm$ 0.03  & 370 $\pm$ 30  &  900 $\pm$ 50 \\
  {0150+13}         &   --  &$< 0.01$&  0.79 $\pm$ 0.02  & 260 $\pm$ 50  &  --           \\
  {0938+54}         &   --  &$< 0.05$&  1.97 $\pm$ 0.04  & 350 $\pm$ 30  & 550 $\pm$ 100 \\
  {2103$-$07}       &   --  &$< 0.01$&  0.52 $\pm$ 0.03  & 130 $\pm$ 20  & 700 $\pm$ 50 \\
 \hline
\end{tabular}
\tablefoottext{1}{from \citet{Bergvall06,Heckman11,Leitet13} 
plus Tol\,1214$-$277\citep{Thuan97,Mas-Hesse03}, 
and GP1291+15 \citep{Cardamone09}, for which medium-resolution \lya\ spectra 
are available. Below the horizontal line we also list four 
non-leakers from the sample of \citet{Heckman11} for comparison.}
\tablefoottext{$\ast$}{Official catalogue names: Tol\,1247$-$232, Haro\,11, 
       Tol\,1214$-$277,    
       \object{SDSSJ121903.98+152608.5},
       \object{SDSSJ021348.53+125951.4}, \object{SDSSJ080844.26+394852.3},
       \object{SDSSJ092159.39+450912.3}, \object{SDSSJ092600.40+442736.1},
       \object{SDSSJ005527.46$-$002148.7}, \object{SDSSJ015028.40+130858.3}, 
       \object{SDSSJ093813.49+542825.0}, \object{SDSSJ210358.74$-$072802.4}
       }
\tablefoottext{a}{From \citet{Leitet13}. }
\tablefoottext{b}{Taken from \citet{Heckman11} if not indicated otherwise.} 
\tablefoottext{c}{\oiiil/\oiill\ derived from SDSS data for \citet{Heckman11} sample and for
the green peas; from \citet{Terlevich91} for Tol\,1214$-277$ and 
Tol\,1247$-$232; and from \citet{Guseva12} for Haro\,11.}
\tablefoottext{d}{Measured main \lya\ peak offset. Systemic emission-line redshifts 
adopted from the SDSS for the sample of \citet{Heckman11} and for the green peas; 
from the 6dF survey of Haro\,11; from \citet{Pustilnik07} for Tol\,1214$-$277; 
from the H$\alpha$ line (E.\,Freeland, private communication) for Tol1247-232.}
\tablefoottext{e}{Based on \lya\ in this study; cf.\ also \citet{Mas-Hesse03} .}
\tablefoottext{f}{\citet{Thuan97}}
\end{table*}

To compare our theoretical prediction with observations,
we now examine low-redshift objects from the literature
for which direct Lyman continuum escape has been detected or which
are considered candidates for LyC leakage on the basis of indirect indicators.
Because possibilities for directly probing the ionising continuum in nearby galaxies 
have been limited to the archival FUSE data,
indirect methods have been used frequently.  
These mainly include the measurement of saturation
in the UV absorption lines of metals in their low-ionisation states,
which form in the \hi\ regions. Residual flux in the lines indicates
low optical depth or incomplete covering of the UV-light source
\citep[e.g.][]{Heckman01,Jones13}.                       
Independently, the \oiiill/\oiill\ line ratio has been proposed as
a LyC optical-depth indicator \citep{Jaskot13,Nakajima14}.
In density-bounded \hii\ regions, the \oiii\ excitation
zones are expected to be largely unaffected, whereas the
outer \oii\ zones will be truncated, resulting in high
\oiii/\oii\ ratios \citep{Kewley13}.

Our sample was selected from \citet{Bergvall06, Heckman11, Leitet13}. 
To this, we added Tol\,1214$-$277 and GP1219+15,  
which show a \lya\ profile that is highly suggestive of Lyman-continuum
leakage according to our predictions in Sects.\ \ref{s_riddled} and 
\ref{s_densitybounded}.
The low-redshift data are convenient for this study because of the availability of 
high-resolution \lya\ spectra and the rich ancillary data. 
\lya\ spectral properties of the sample are summarised in Table \ref{Tab}, 
together with the other criteria used for identification of possible 
LyC leakage.  
Column 2 indicates \fesclyc measured by direct Lyman-continuum-emission 
detection; 
Col.\ 3 gives \fesclyc deduced from the UV covering fraction
estimated from the LIS absorption lines;
Col.\ 4 lists the observed \oiiil/\oiill\ ratio, 
Col.\ 5 gives \lya\ peak offset, and Col.\ 6 shows the peak separation for 
double-peaked profiles.

\subsection{Redshift and \lya\ peak measurements}
We present the UV \lya\ line profiles of 11 objects in Figs.~\ref{lya_leakers}, 
~\ref{CF_leakers}, and ~\ref{figGP} (Table\,\ref{Tab} contains 12 entries, 
but the spectrum of Tol\,1247$-$232 will be part of 
a detailed study of Puschnig et al. in prep). 
The data were mostly observed with HST/COS, one object
was observed with HST/GHRS (Tol1214$-$277), and they were retrieved from the MAST multi-mission catalogue.
We box-car smoothed the archival data to the 
nominal resolution of each of the spectrographs 
(COS $R=20\,000$ for a point source,
GHRS $R=2000$).  
However, our sources are not point-like, and we  
estimate the real spectral resolution of COS \lya\ to be 
$\sim$\,100\,\kms, or $R \sim 3000$ (Orlitova et al. in prep).
We measured the main red \lya\ peak offset (\vpeak) from the systemic 
velocity (Col.\ 5 of Tab.\ \ref{Tab}) and the separation $S_{\rm peak}$ of 
the main red and blue \lya\ peaks of the double-peaked profiles (Col.\ 6 of 
Table\,\ref{Tab}). 
The main peak was defined as the flux maximum in each 
case (with no fitting performed). We comment on the cases where other 
significant peaks are present, especially if they are located
closer to the systemic redshift,
in the description of individual objects.
Accurate systemic redshifts are necessary to reliably determine 
\vpeak.  
We assume that the intrinsic \lya\ is formed in the same \hii\ regions
and by the same mechanisms as H$\alpha$ and other Balmer lines, 
and therefore redshifts derived from the
optical emission lines are relevant. Where available, we applied
SDSS emission-line redshifts. For objects outside the 
SDSS database (Haro\,11, Tol\,1214$-$277 and Tol\,1247$-$232), we adopted emission-line 
redshifts with the smallest error bars available through the NED database
(see references in Table\,\ref{Tab} and Fig.\,\ref{lya_leakers}).       
The uncertainties in the measured peak positions are mainly due to the redshift
errors, the noise, and the peak irregular shapes. We checked that  
errors in wavelength calibration contribute less than 10\,\kms.

\subsection{Haro\,11 and Tol\,1247$-$232}

\object{Haro\,11} is a known LyC leaker with $f_{\rm esc}(LyC) \sim 3\%$ 
measured in the FUSE spectra \citep{Bergvall06,Leitet13}.
It is a blue compact galaxy, and its optical, UV, and \lya\ morphologies
 were studied with HST in \citet{Kunth03, Hayes07}, and \citet{Ostlin09}.
 Knot C was found to have \lya/\ha~$\sim 10,$ indicating a direct \lya\
 escape from this region through holes or an optically thin medium. 
The galaxy is composed of three major condensations (knots).  
Cold gas observations through the \ion{Na}{i} D line by \citet{Sandberg13}
suggest that the ISM is clumpy.      
The \lya\ profile of knot C has recently been acquired with HST/COS
(GO 13017, PI: T.Heckman, top panel of Fig.~\ref{lya_leakers}).
It presents a standard asymmetric, redshifted line with \vpeak\ = $115 \pm 50$\,\kms. 
It is therefore just compatible with our prediction of LyC escape from a 
uniform 
medium of low column density, which is indicated by $\vexp \la 150$\,\kms.
Its \oiii/\oii\ $\sim 1.4$ \citep{Guseva12} ratio is moderate,
as is its LyC escape fraction \fesclyc = 0.032 derived from direct LyC measurement, 
and \fesclyc = 0.024 derived from the LIS absorption lines \citep{Leitet13}.

\object{Tol\,1247$-$232}, the second-best candidate for LyC leakage, 
with a direct detection of non-zero Lyman continuum with FUSE \citep{Leitet13}, 
has recently been observed with HST/COS (GO 13027, PI: G.\"Ostlin, Puschnig et al. in prep). 
The \lya\ profile is similar to that of Haro 11, 
but broader, and peaks around \vpeak$= 150 \pm 70$\,\kms, (marginally) compatible 
with \vpeak $\leq 150$\,\kms. Its \oiii/\oii\,$\sim3.4$  is the third-highest 
value of our sample. 
The covering fraction of its LIS absorption lines remains to be determined from 
the new COS observations.

\subsection{Tol\,1214$-$277}

\object{Tol\,1214$-$277} is a low-metallicity, compact blue dwarf galaxy 
\citep[][]{Thuan97}.
It was unfortunately not observed with FUSE.
However, its \lya\ emission profile, observed with HST/GHRS
\citep[bottom panel of Fig.~\ref{lya_leakers}, see also][]{Thuan97,Mas-Hesse03},
peaks at the systemic velocity, it is thus a nice example of \vpeak$\sim 0$.
The \lya\ profile seems to be composed of three sub-peaks, similar to  
the spectrum emerging from a static medium with a covering fraction CF=0.90 
presented in the left panel of Fig.~\ref{ionisation_bounded}
\citep[see also the triple-peaked profiles predicted by][]{Behrens14}. 
A spectrum with higher resolution would be needed for a better
understanding of the profile. However, the important feature is the profile
centred on the systemic redshift, which cannot be reproduced by any model 
out of our grid of $>6000$ radiation transfer models for homogeneous shells  
\citep{Schaerer11}.
We can reproduce the external peaks alone with a static homogeneous model.
The central peak can either result from a high-velocity component or from  
clumpy (riddled) medium with incomplete covering, which allows for LyC 
leakage (as in Fig.~\ref{ionisation_bounded}).

The \ion{C}{II}\,$\lambda1335$ absorption line shows two velocity
components, one centred on the systemic velocity, the other blueshifted by 
950\,\kms.
The rapid motions can contribute to the \lya\ escape at systemic velocity. 
However, both \ion{C}{II} components are unsaturated, which may
indicate a non-uniform 
coverage of neutral gas along the line of sight, which would favour our scenario of 
LyC escape through holes.    
The same qualitative conclusion was reached by \citet{Thuan97}. 

An alternative interpretation was provided by \cite{Mas-Hesse03}. 
They suggested that the central component corresponds to direct \lya\ emission
from an ionised cavity, while the two additional emission peaks originate from  
ionisation fronts trapped within the expanding shell (both approaching and receding).
In this case, the outflow velocity is predicted to be traced by the lateral peaks at
$\vexp \approx 300$\,\kms.

The observed ratio of \oiii/\oii\ $\sim 15$ of Tol\,1214$-$277 is very high,
suggesting high ionisation (and/or low metallicity) of the \hii\ region
\citep[][]{Kewley02,Kewley13,Jaskot13,Nakajima14}.

\subsection{Lyman-break analogues}

\citet{Heckman11} have identified three LyC-leaking candidates among
their eight Lyman-break analogues (LBAs) that were observed with HST/COS. 
They defined the LyC-leaking candidates as those that show incomplete
covering of the far-UV source by neutral gas clouds (derived from 
the UV LIS absorption lines), and whose UV morphology is dominated by 
a central compact source. 
They did not consider 0926+44 a leaking candidate although 
it has the lowest UV-covering fraction of the entire sample as
a result of the 
absence of the dominant central core.   
We here consider this fourth object together 
with the other three leaking candidates and show their \lya\ profiles 
in the left panel of Fig.~\ref{CF_leakers}.
The remaining four non-leaking galaxies from their sample are shown in 
the right panel of Fig.~\ref{CF_leakers}. 

We find that all of the objects identified as potential leakers have
unusual rather complicated \lya\ spectra with bumps and substructures in 
the main peaks that are impossible to model completely with radiative transfer 
models in homogeneous expanding shells. 
In contrast, the non-leaking objects have classical P Cygni-type \lya\ profiles.
Two leaking candidates (0213+12 and 0808+39) have 
a very low \vpeak\ ($\sim70$\,\kms and $\sim30$\,\kms, 
respectively, see Table\,\ref{Tab}) and in addition show a subpeak
at the systemic redshift, which may indicate \lya\ 
transfer through a riddled ISM. The subpeak in 0213+12 is weak, 
but the main double-peak structure is narrow and the object 
could also be classified as a LyC-leaking candidate with a low column density.
The \lya\ profile of 0926+44 has a maximum at $280\pm90$\,\kms
and a blue peak separated by $\sim450$\,\kms, which excludes LyC leakage.
However, a strong subpeak is present at $100\pm90$\,\kms, which fits
our LyC leaking criteria. The \lya\ spectrum thus seems to be composed of
two components that can either be due to complex morphology and kinematics
of the galaxy \citep{Basu-Zych09,Goncalves10} or to high-velocity outflows
\citep{Heckman11} and is not straightforward to interpret.

Three of the four leaking candidates reported by Heckman emit significant flux bluewards of the systemic redshift, 
as pointed out by \citet{Heckman11}, but the separation between the peaks 
is too high to fit our predictions of leakage through the optically thin scenario, 
except for 0213+12. 
The object 0921+45 of the leaking canditates listed by Heckman et al. does not meet 
our criteria for LyC leakage because of its large \lya\ absorption trough 
around the systemic redshift, indicating large amounts of neutral gas and dust. 
It has recently been observed in the LyC continuum, and its absolute escape fraction is 
indeed low ($\sim 1\%$), although the relative escape is higher 
\citep[$\sim 21\%$][]{Borthakur14}.
Of the non-leakers, three out of four have \vpeak\ $\gg$ 150\,\kms\ and no 
\lya\ escape bluewards of the systemic redshift. Indeed, these objects
present a broad absorption bluewards of the systemic redshift. 
As an exception among the non-leakers of Heckman, 2103$-$07 has 
\vpeak\,$\leq$\,150\,\kms and its \lya\ red peak has a substructure that
could also be interpreted as a subpeak at $v\sim50$\,\kms. However, the 
LIS absorption lines are saturated and indicate no LyC escape. 
In addition, its large \lya\ absorption trough suggests a large amount of dust.  
The low \vpeak\ does not contradict our LyC prediction criteria
because low \vpeak\ values can also appear in high column density objects
(see Fig.\,\ref{vpeak}), and the \vpeak\ alone is not sufficient to predict the LyC leaking. 
         
The \oiii/\oii\ ratios are all below $\sim3$  for both leakers and non-leakers, 
and the mean \oiii/\oii\ values are almost identical ($<$\oiii/\oii$> = 1.22$ for 
leakers and $1.20$ for non-leakers).
None of the objects in the sample of Heckman have the extreme \oiii/\oii\ ratios reported in 
\citet{Jaskot13} to potentially trace density bounded \hii\ regions.

\subsection{Other leaking candidates in the literature}
Among the LyC-leaking candidates presented in \citet{Leitet13}, 
\object{ESO\,338$-$IG04} and \object{SBS\,0335$-$052} are the two with the strongest
continuum escape fractions derived from the UV LIS lines. 
They were both observed with FUSE. Unfortunately, 
they are so close in redshift that it is impossible to constrain 
the LyC escape from these observations.
Their \oiii/\oii\ ratios are $\sim$\,4 and
$\sim$\,12, respectively \citep{Guseva12,Izotov09}. However, SBS\,0335$-$052 is a 
\lya\ absorber \citep{Thuan97}, and therefore our \lya\ emission profile diagnostics
are not applicable. It is the second-most metal-poor blue compact dwarf known to date,  
and therefore its \oiii/\oii\ ratio probably results from the extremely low metallicity
\citep[see e.g.][for a discussion of this effect]{Kewley13}.
It is embedded in an enormous \hi\ cloud, and thus we believe that SBS\,0335$-$052 is not a 
good leaking cadidate.

For ESO\,338$-$IG04, we have a UV slit spectrum observed with HST/STIS, but unfortunately 
in the low-resolution mode ($R=1000$). The analysis of spectra from the emission knots   
\citep{Hayes05} shows great spatial variability, ranging from absorption to emission. 
The \lya\ spectrum that emerges from knot D shows $\vpeak \sim 0$ and therefore seems
compatible with LyC leakage according to our \lya\ diagnostics presented in this study. 
However, a high-resolution spectrum is necessary for a confirmation.

\begin{figure}
  \includegraphics[width = 0.5\textwidth]{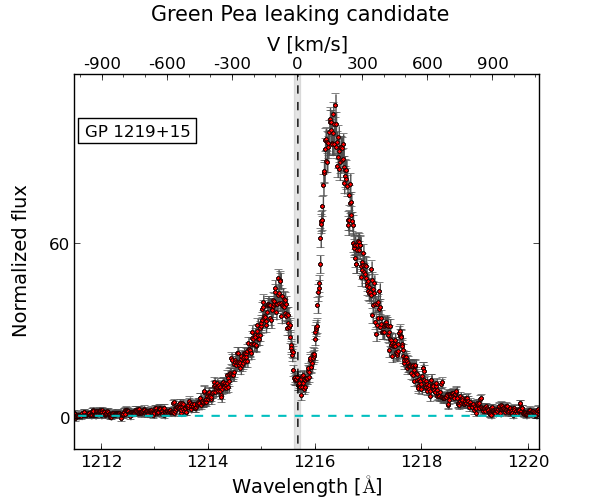} 
  \caption{Green pea 1219+15 is a good candidate for LyC leaking according to 
our \lya\ diagnostics because the double peaks are separated by $\sim 270 < 300$\,\kms.}
  \label{figGP}
\end{figure}

Recently, UV spectra of green peas \citep{Cardamone09} acquired with HST/COS
have become available through the MAST archive 
\citep[GO 12928, PI: A.Henry, GO 13293, PI: A.Jaskot,][]{Jaskot14}. 
We have analysed several of their \lya\ profiles and are working on modelling them 
with the radiation transfer code \McLya\ (Orlitov\'a et al. in prep).
We find that the object with the highest \oiii/\oii\ ratio ($\sim$\,10), GP\,1219+1526,
has a narrow double-peak \lya\ profile ($S_{\rm peak}\sim\,270\, < 300$\,\kms,
 $\vpeak \sim 170 \pm 20$\,\kms, see Fig.\,\ref{figGP}), 
indicating probable LyC leaking from this object.

A unique test of our LyC-\lya\ connection is possible
thanks to the active galactic nuclei (AGN), where
strong flux of ionising continuum is known to escape
from type 1 sources, while it is missing in type 2 AGN.
We therefore expect a difference between \lya\
spectra in the two types.
And indeed, in agreement with our predictions,
\lya\ spectra of Seyfert\,1 galaxies
and type 1 quasars in low and high redshift show
strong \lya\ emission lines at systemic velocity
\citep{Gregory82,Rudy88,Crenshaw99,Kraemer01,Shull12,Herenz15}.
We examined the spectra directly in the MAST archive, where
possible. These sources include types 1\,--\,1.8, which means that they have a partially
obscured AGN, in line with our models. 
NGC\,4303, the only true type 2 Seyfert galaxy
for which we have found a \lya\ spectrum, has a P Cygni profile with a
redshifted peak \citep{Colina02}.

\section{Discussion}
\label{s_discuss} 

In the data available with low redshift, we have seen that it
is rare to find spectra 
with the \lya\ peak centred on systemic redshift, suggesting a clumpy ISM. Instead, our criteria related to the low column
density neutral ISM 
were used more frequently. 
We found that the two objects with directly detected LyC, that
is, Haro\,11 and
Tol\,1247$-$232 \citep{Leitet13}, indeed have \lya\ spectra characterised by 
\vpeak$<150$\,\kms, as predicted by our models. Therefore, the \lya\ spectral
diagnostics do not contradict the LyC measurements.

Among the candidates identified by weak LIS absorption lines, 
we found \lya\ spectra with interesting features, 
and at least some of them would be classified as LyC-leakage candidates according to 
our criteria (Tol\,1214$-$277, LBA 0213+12, LBA 0808+39, GP 1219+15).  
Consistently, the objects where LyC leakage was excluded by the LIS absorption
lines method show \lya\ spectra that clearly result from high column 
densities of neutral gas and/or dust. 
The cases in which weak LIS absorption lines were detected, but where at the same
time our criteria exclude LyC leakage, have unusual \lya\ spectra 
(LBA 0921+45, LBA 0926+44), and seem to have complex galaxy morphologies 
\citep{Goncalves10} and would need a more detailed study.

Therefore, based on the available data sample, we are confident that 
our theoretical prediction of \lya\ features indicating LyC leakage   
is a viable approach for identifying leaking candidates
at both low and high redshift.

We have also described possible limitations of our method:
1) relatively high spectral resolution ($R>4000$) is required for a clear 
interpretation of the spectral profile; 2) redshift precision is crucial 
for correctly determining the \lya\ peak offset; 3) one of our criteria 
relies on double-peaked \lya\ profiles, which are rare; 
4) our method only indicates leaking candidates and 
cannot alone prove the LyC leakage. Detailed modelling of the individual \lya\
spectra will help to distinguish between the cases of \lya\ peaks close to the
systemic redshift that are due to the ISM that is optically thin for LyC and
those that are only transparent to \lya\ as a result of fast outflows or other conditions.          
However, the ultimate confirmation of LyC leakage for every selected candidate
must come from the direct LyC detection. 
We have shown that our \lya\ spectral criteria can reliably identify non-leakers 
in which LyC leakage is excluded.

\citet{Erb10,Erb14,Heckman11} concluded based on observational results 
that \lya\ flux bluewards of the systemic redshift is related to the LyC leakage. 
This is similar to our criteria in the
case where the neutral ISM is clumpy and the observed \lya\ profile is centred
on the systemic redshift. However, the blue-flux objects also contain those  
that have double peaks arising from a static ISM. 
Our method provides a theoretical background to these observational findings 
and sets clear criteria for LyC signatures in the \lya\ profiles.

A comparison with \oiii/\oii\ is less straightforward.
Galaxies with \oiii/\oii\ $< \sim 4$ are characterised by a wide spread
in \vpeak\ and \speak, whereas only a few objects have larger \oiii/\oii.
The wide spread is consistent with results reported by \citet{Nakajima14}, who showed that
\oiii/\oii\ as a LyC diagnostics is strongly affected by the degeneracy
between metallicity and ionisation parameter.
Large statistical galaxy samples are needed to test the
correlation between \oiii/\oii\ and the \lya\ profile parameters. 
See also \citet{Stasinska15}.

Obviously, larger galaxy samples with both high-resolution \lya\ spectra 
and measured $f_{\rm esc}(LyC)$ are needed to derive statistically 
robust conclusions. However, we believe that our theoretical predictions
of Sect.\,\ref{s_LyaLyC} (i.e., low \vpeak\ and $S_{\rm peak}$, or \lya\ peak
at systemic redshift for LyC leaking candidates) 
have been shown to be useful based on the presented HST archive sample. 
Several HST/COS programs that investigate the Lyman-alpha spectra and
LyC escape fraction from local galaxies are ongoing.
By the end of 2015, the number of HST/COS \lya\ spectra available 
will be $> 100$, which will provide an important sample size, and hopefully,
new insights into the LyC-leakage problem will be gained.

\section{Conclusions}
\label{s_conclude}

We have carried out  \lya\ radiation transfer simulations in two idealised
configurations of LyC-leaking star-forming galaxies:
\begin{enumerate}
\item Homogeneous spherically expanding ISM shells with a central \lya\ source, 
but with extremely low column densities ($N_{\rm HI} \leq 10^{18}$ cm$^{-2}$),
\item clumpy spherical shells with a non-unity covering fraction, 
called riddled ionisation-bounded \hii\ regions.
\end{enumerate}
In both cases, we find that the emerging \lya\ spectrum has remarkable features
that may be possible to identify in observed \lya\ spectra, assuming that 
the \lya\ spectrum from a galaxy may be dominated by the \lya\ component 
emerging from LyC-leaking star clusters. 
According to the first scenario, the \lya\ spectrum will have a classical asymmetric
redshifted profile, but with a small shift (\vpeak\ $\leq 150$\,\kms).
According to the second scenario, the main feature is a peak at the systemic redshift 
of the star cluster, with, as a consequence, a non-zero flux bluewards of the \lya\ 
line centre. These two signatures may be used to distinguish
or select leaking 
candidates at all redshifts for objects with high spectral resolution 
and well-determined systemic redshift.

We have examined how these diagnostics compare with observed \lya\ spectra 
from the local Universe where either the LyC escape fraction has been directly 
measured \citep{Bergvall06, Leitet13} or has been derived from LIS absorption 
lines studies \citep{Heckman11}. We also tested our predictions against a sample of non -leaking galaxies of the same team. Although the number of objects is very small
to derive statistically robust conclusions, the expected trends seem to be at work 
(\vpeak\ and $S_{\rm peak}$ smaller for leakers than non-leakers).

We proposed two additional galaxies as good candidates for LyC leaking: 
GP\,1219+1526, based on the small separation of the peaks in its \lya\ profile,
and a high \oiii/\oii\ ratio, could be leaking according to our first scenario of 
a density-bounded region;
Tol\,1214--277, described in \citet{Thuan97}, based on its very 
peculiar \lya\ spectral shape (symmetrical triple peak with the main peak at 
\vpeak\ = 0), and indications of non-total coverage of LIS absorption lines,
could be leaking from a riddled ISM. 

There are well-known objects in the local Universe from which a strong LyC flux is 
detected, belonging to the family of AGNs. As another validation of our diagnostics, 
we verified in the literature that \lya\ spectra from these known LyC leakers are in 
emission, and peak at the line centre.

Our predictions need to be tested on larger samples with well-determined 
systemic redshifts and high spectral resolution \lya\ profiles. 
But apparently a small shift of the maximum of the \lya\ profile 
with respect to the systemic redshift (\vpeak\ $\leq 150$\,\kms) 
may be a good proxy for a non-zero escape fraction of ionising radiation from galaxies.


\begin{acknowledgements} 
  We thank Florent Duval for sharing radiation 
  transfer and clumpy geometry models from Duval et al. (2014).
  We are grateful to the anonymous referee and to Miguel Mas Hesse for their
  thoughtful comments that helped us improve the paper.
  A.V. is supported by a Fellowship "Boursi\`ere d'Excellence" of 
  Geneva University. I.O. acknowledges the support from the 
  Sciex fellowship of the 
  Rectors' Conference of Swiss Universities, and the grant 14-20666P 
  of Czech Science Foundation, together with the long-term institutional 
  grant RVO:67985815.
  M.H. acknowledges the support of the Swedish Research Council, 
  Vetenskapsr{\aa}det, and the Swedish National Space Board (SNSB).
  This work used NASA/ESA HST archival data from programmes
  GO 6678, GO 9036, GO 11727, GO 12928, GO 13017, and GO 13027.
  All of the data presented in this paper were obtained from 
  Mikulski Archive for Space Telescopes (MAST). STScI is operated by
  the Association of Universities for Research in Astronomy, Inc.,
  under NASA contract NAS5-26555. Support for MAST for non-HST data is
  provided by the NASA Office of Space Science via grant NNX13AC07G and
  by other grants and contracts.
  We made use of the SAO/NASA Astrophysics Data System (ADS),
  the NASA/IPAC Extragalactic Database (NED), and the Image Reduction
  and Analysis Facility (IRAF), distributed by the
  National Optical Astronomy Observatories.

\end{acknowledgements}


\bibliography{references}

\end{document}